\documentclass[aps,showpacs,11pt]{revtex4}
\usepackage{dcolumn}
\usepackage{graphicx}
\usepackage{amsmath}
\usepackage{amsfonts}
\usepackage{amssymb}
\usepackage{psfrag}
\usepackage{wrapfig}
\usepackage{subfigure}
\usepackage{makeidx}
\usepackage{bm}
\usepackage{epsf}
\usepackage{hyperref}
\usepackage{color}
\usepackage{float}
\usepackage{epsfig}
\usepackage{epstopdf}

\begin{document}

\title{Superradiant instability of Kerr-de Sitter black holes in scalar-tensor theory}

\author{Cheng-Yong Zhang, Shao-Jun Zhang and Bin Wang}

\affiliation{Department of Physics and Astronomy,
Shanghai Jiao Tong University, Shanghai 200240,
China}

\date{\today}

%\keywords{superradiance, classical stability, scalar-tensor, black hole}

\begin{abstract}
\indent

We investigate in detail the mechanism of
superradiance to render the instability of
Kerr-de Sitter black holes in scalar-tensor
gravity. Our results provide more clues to
examine the scalar-tensor gravity in the
astrophysical black holes in the universe with
cosmological constant. We also discuss the
spontaneous scalarization in the de Sitter
background and find that this instability can also
happen in the spherical de Sitter configuration in a special style.

\end{abstract}
\maketitle

\section{Introduction}

Scalar-tensor (ST) theory is an alternative theory
that describes gravity beyond Einstein's general
relativity (GR). This theory was originally
conceived by Jordan and later generalized by
Brans and Dicke \cite{BD}.  The primary motivation
comes from cosmology to search for a theory
incorporating Mach's principle which is not
explicitly embodied in GR. In the ST
theory, the gravitational coupling is determined
by all matter in the universe. The cosmological
distribution of matter will affect local
gravitational experiments \cite{STBook}. Currently, interests in ST theory come
from many aspects. First, a fundamental scalar
field coupled to gravity is an unavoidable
feature of unified theory such as supergravity,
string theory \cite{Green:1987mn}, etc. Second, the
inflationary scenarios of universe and dark
energy may be described by ST theory
\cite{STBook}.

Experimentally, the differences between
ST theory and GR are too small to be detected in the weak
field regime, i.e, in the solar system \cite{test}. Particular interest of the
phenomenology of the ST theory is in
the strong gravity regime. It is expected that
the strong gravity test can be a possible way to
distinguish the ST theory from GR.

For compact stars in ST theory,
unexpected phenomenon has been discovered
compared with GR \cite{hair}. For the black hole (BH)
background in ST theory, when the
scalar field settles to a constant, we can get
the same BH solution as that in GR
\cite{Uniq}. However the perturbations will
behave differently as the two theories have
different dynamics \cite{different}. It has been
shown that compared with the Kerr BH
background in GR, the existence of a scalar mode
in the spectrum of perturbations around a Kerr
BH in the ST theory leads to
remarkable effects \cite{Cardoso}. When the
matter configuration surrounding the Kerr black
hole is dense enough, the BH will be
forced to develop scalar hair. On the other hand,
when the BH rotates sufficiently fast,
superradiant instability can occur. Neither of
these instabilities has been observed in the Kerr
BH solution background in GR. This gives
the hope that BHs can be used as probes
in order to distinguish ST theory from
GR.

Considering that our universe has a positive
cosmological constant \cite{Weinberg:2008zzc},  we will
extend the discussions in the asymptotically flat
Kerr BH backgrounds \cite{Cardoso} to
Kerr-de Sitter configurations. We will examine in
detail the mechanism that can render the
instability of Kerr-de Sitter BHs driven
by superradiance in the ST gravity. To
trigger superrandiant instability, two necessary conditions need to be satisfied \cite{Hod2013}, namely
that the BH must have superradiance and
the existence of a potential well outside the
BH to trap the scattered wave. The
superradiant instability in GR has been discussed
thoroughly, see review \cite{Cardoso2013} and
references therein. In this work, we will examine
how the instability depends on the matter
profile, the rotation of the BH and the
cosmological constant. Furthermore we will
investigate the spontaneous scalarization in the
Kerr-de Sitter BH background. Comparing
with the stability analysis in GR \cite{QNM of
Kerr-dS} where no instability was found for
Kerr-de Sitter BH configurations, our
results will provide more possible potential
observational clues to distinguish the
ST gravity from GR in the BH
backgrounds in the real universe.

The organization of the paper is as follows. In
section II, we will briefly present the framework
of ST theory and describe the Kerr-de
Sitter BH. In section III, we will derive
the perturbation equation of the scalar field. In
section IV, we will calculate the frequencies of
the perturbation numerically and list our
results. In section V, we will discuss the
spontaneous scalarization for spherical de Sitter
BHs. The last section is devoted to
discussion and summary.

\section{Kerr-de Sitter black hole in scalar-tensor theory}

The general action of the ST theory in
the Jordan frame is \cite{STBook}
\begin{eqnarray}
S & = & \frac{1}{16\pi}\int d^{4}x\sqrt{-g}\left(F(\phi)R^J-Z(\phi)g_{\mu\nu}\partial^{\mu}\phi\partial^{\nu}\phi-U(\phi)\right)\nonumber \\
 &  & +S_{m}(\Psi_{m};g_{\mu\nu}),
\end{eqnarray}
where $R^J$ is the Ricci scalar of metric
$g_{\mu\nu}$, $\phi$ is a scalar field and
$U(\phi)$ is the scalar field potential.
$S_m(\Psi_{m};g_{\mu\nu})$ is the action describing
matter $\Psi_{m}$ which is minimally coupled to
$g_{\mu\nu}$ and $\phi$. Functions $F$, $Z$ and
$U$ represent specific theories within the class,
up to a degeneracy due to the freedom of
redefining the scalar \cite{redefine-scalar}.
When $F=\phi$, $Z=\omega_{0}/\phi$ and $U=0$, one
gets the Brans-Dicke theory \cite{BD}. The ST theory can be viewed as a low-energy limit
of a bosonic string theory when $F=\phi$,
$Z=-\phi^{-1}$.

Performing a conformal transformation and the
field redefinition
\begin{eqnarray}
g_{\mu\nu}^{E} & = & F(\phi)g_{\mu\nu},\nonumber \\
\Phi(\phi) & = & \frac{1}{\sqrt{4\pi}}\int d\phi\left(\frac{3}{4}\frac{F'(\phi)^{2}}{F(\phi)^{2}}+\frac{1}{2}\frac{Z(\phi)}{F(\phi)}\right)^{1/2},\\
A(\Phi) & = & F^{-1/2}(\phi),\nonumber \\
V(\Phi) & = & \frac{U(\phi)}{F^{2}(\phi)},\nonumber
\end{eqnarray}
we get the action in the Einstein frame (For more
details, see \cite{STBook})
\begin{eqnarray}
S & = & \frac{1}{16\pi}\int d^{4}x\sqrt{-g^{E}}\left(R^{E}-8\pi g_{\mu\nu}^{E}\partial^{\mu}\Phi\partial^{\nu}\Phi-V(\Phi)\right)\nonumber \\
 &  & +S_{m}(\Psi_{m};A(\Phi)^{2}g_{\mu\nu}^{E}).\label{eq:action-Einstein}
\end{eqnarray}
In Einstein frame, the scalar field is minimally
coupled to gravity, but the matter field is
coupled to the metric
$A(\Phi)^{2}g_{\mu\nu}^{E}$. The field equations
in the Einstein frame can be derived by varying
action (\ref{eq:action-Einstein}).
\begin{eqnarray}
G_{\mu\nu}^{E} & = & 8\pi\left(T_{\mu\nu}^{E}+\partial_{\mu}\Phi\partial_{\nu}\Phi-\frac{g_{\mu\nu}^{E}}{2}(\partial\Phi)^{2}\right)-\frac{g_{\mu\nu}^{E}}{2}V(\Phi),\\
\nabla_{\mu}\nabla^{\mu}\Phi & = & -\frac{A'(\Phi)}{A(\Phi)}T^{E}+\frac{V'(\Phi)}{16\pi}.
\end{eqnarray}
Here $T_{\mu\nu}^{E} \equiv -\frac{2}{\sqrt{-g^{E}}}\frac{\delta S_{m}}{\delta g_{E}^{\mu\nu}}$.
We assume that $\Phi=\Phi_{0}=const$ is a solution to the above equations and around $\Phi_0$ we
have the following analytical expansions
\begin{eqnarray}
V(\Phi) & = & \underset{n=0}{\sum}V_{n}(\Phi-\Phi_{0})^{n},\\
A(\Phi) & = &
\underset{n=0}{\sum}A_{n}(\Phi-\Phi_{0})^{n}.
\end{eqnarray}
For abbreviations, we denote $\varphi \equiv
\Phi-\Phi_{0}$ and $\alpha_{n} \equiv
A_{n}/A_{0}$. Then to the order
$\mathcal{O}(\varphi)$, the equations of motion
become
\begin{eqnarray}
G_{\mu\nu}^{E}+\frac{V_{0}}{2}g_{\mu\nu}^{E} & = & 8\pi T_{\mu\nu}^{E}-\frac{g_{\mu\nu}}{2}V_{1}\varphi,\\
\nabla_{\mu}\nabla^{\mu}\varphi & = & -\alpha_{1}T^{m}+\frac{V_{1}}{16\pi}+(\alpha_{1}^{2}-2\alpha_{2})T^{m}\varphi+\frac{V_{2}}{8\pi}\varphi.\label{eq:ScalarEq}
\end{eqnarray}
The term $V_{0}$ is related to the cosmological
constant $\Lambda\equiv\frac{V_{0}}{2}$. $V_{1}$
is a linear potential which will be dropped off
since linear potential is seldom met in nature.
$V_{2}$ is related to a standard mass term.
$\alpha_1$ describes the effective coupling
between the scalar and matter. To have a constant
scalar solution $\Phi=\Phi_{0}=const$ as we
mentioned above, from ($\ref{eq:ScalarEq}$) we
should set $\alpha_1=0$. In addition,
observations, such as weak gravity constraints
and tests for violation of the strong equivalence
principle, seem to require $\alpha_{1}$ to be
negligibly small \cite{alpha3}. The configuration
with constant scalar and $\alpha_{1}\simeq 0$ was
argued most likely as an approximate solution in
most viable ST theories
\cite{Cardoso}. Finally, the equations of motion
can be reduced to
\begin{eqnarray}
G_{\mu\nu}^{E}+\Lambda g_{\mu\nu}^{E} & = & 8\pi T_{\mu\nu}^{E},\label{eq:GravityEq}\\
\nabla_{\mu}\nabla^{\mu}\varphi & = & \left(\frac{V_{2}}{8\pi}-2\alpha_{2}T^{m}\right)\varphi.\label{eq:ScalarEq-1}
\end{eqnarray}
Here
$\mu_{s}^{2}\equiv\frac{V_{2}}{8\pi}-2\alpha_{2}T^{m}$
plays the role of effective mass square of the
scalar field. When the backreaction of the matter
field on the spacetime geometry is negligible
(``the probe limit"), from (\ref{eq:GravityEq}) we
can have the Kerr-de Sitter BH as a
solution. In Boyer-Lindquist coordinates, the
metric reads
\begin{eqnarray}
ds^{2} & = & -\frac{\Delta_{r}}{(1+\alpha)^{2}\rho^{2}}(dt-a\sin^{2}\theta d\phi)^{2}+\frac{\Delta_{\theta}\sin^{2}\theta}{(1+\alpha)^{2}\rho^{2}}(adt-(a^{2}+r^{2})d\phi)^{2}\nonumber \\
 &  & +\frac{\rho^{2}}{\Delta_{r}}dr^{2}+\frac{\rho^{2}}{\Delta_{\theta}}d\theta^{2},
\end{eqnarray}
in which
\begin{eqnarray*}
\alpha=\frac{\Lambda a^{2}}{3}, & \Delta_{\theta}=1+\alpha\cos^{2}\theta, & \rho^{2}=r^{2}+a^{2}\cos^{2}\theta,
\end{eqnarray*}
\begin{eqnarray}
\Delta_{r} & = & (a^{2}+r^{2})(1-\frac{\alpha}{a^{2}}r^{2})-2Mr.
\end{eqnarray}
For some ranges of parameters, $\Delta_{r}=0$
exists four real roots. Three positive real roots
correspond to three horizons: the inner BH Cauchy horizon $r_-$, the outer BH
event horizon $r_{h}$ and the cosmological
horizon $r_{c}$. The remaining negative root has
no physical meaning. We focus only on cases in
which $\Delta_{r}=0$ has four real roots, so that
we have Kerr-de Sitter BH. In figure 1,
we plot the allowed parameter space. We can see
that values of $(a, \Lambda)$ should lie in the
region between two curves where we fix $M=1$. The
upper nearly horizontal curve corresponds to the
degeneracy of the cosmological horizon and the
BH event horizon $r_h=r_c$, while the
lower nearly vertical curve represents the
degeneracy between the BH event horizon
and the Cauchy horizon $r_h=r_-$. The two curves
intersect at the point $(a, \Lambda) \sim
(1.1009, 0.1777)$, at which $r_-=r_h=r_c$ occurs.

\begin{figure}[H]
\centering
\includegraphics[width=0.46\textwidth]{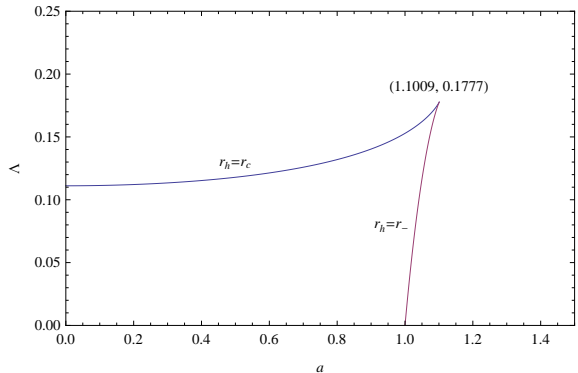}
\caption{\label{fig:a-Lambda}Parameter space for Kerr-de Sitter BH in $a-\Lambda$ plane with fixed $M=1$.
Kerr-de Sitter BH solutions are allowed for the parameters $(a,\Lambda)$ in the region between the two curves.}
\end{figure}

The surface
gravities $\kappa$ and angular velocities $\Omega$ on horizons are \cite{QNM of Kerr-dS}
\begin{eqnarray}
\kappa_{i}=\frac{1}{2(1+\alpha)(r_{i}^{2}+a^{2})}\frac{d\Delta_{r}(r_{i})}{dr} & , & \Omega_{i}=\frac{a}{r_{i}^{2}+a^{2}},
\end{eqnarray}
where $r_{i}=\{r_h, r_c\}$.

In this paper, we will mainly focus on this
Kerr-de Sitter BH configuration in the
probe limit and study its superradiant stability
under scalar perturbation. Different from the
stability analysis in \cite{QNM of Kerr-dS},  we
will look for separable solutions of the
Klein-Gordon equations with effective mass
$\mu_s$, not just the canonical mass of a massive
scalar field. The effective mass reflects the
influence of the ST theory in the
presence of matter surrounding the BH.

\section{Perturbation equations and superradiance}

\subsection{Perturbation equations}

The dynamics of the scalar field in Kerr-de
Sitter background is described by
(\ref{eq:ScalarEq-1}). To make comparison with
the result of Kerr BH in the
asymptotically flat spacetime reported in
\cite{Cardoso}, we take the following form for
the effective mass
\begin{eqnarray}
\mu_{s}^{2}(r)=\frac{2\Lambda}{3}+\frac{G(r)}{\rho^{2}} & , & G(r)=\beta\Theta(r-r_{0})(r-r_{0})\left(\frac{1}{r^{n}}-\frac{1}{r_{c}^{n}}\right),\label{eq:effective-mass}
\end{eqnarray}
which can reduce to the form in Kerr case
\cite{Cardoso} when the cosmological constant
$\Lambda \rightarrow 0$. Here $\Theta(r-r_{0})$
is a heaviside function, $r_{0}$ is the place
where the matter distribution starts from and
$\beta$ indicates the strength of coupling
between scalar and matter field. Note that
$\beta\propto\alpha_{2}$ in (\ref{eq:ScalarEq-1})
and large positive values of $\alpha_{2}$ are not
constrained by observations, so we can take
arbitrary large $\beta$. In fact, we will find
that only large enough $\beta$ can trigger the
instability in the next section. The term
$\frac{2\Lambda}{3}$ plays the role of the
canonical mass term of a massive scalar. For
Kerr-de Sitter background, it equals to
$\frac{1}{6}R_{g}$ where $R_{g}$ is the Ricci
scalar of spacetime. So Eq.(\ref{eq:ScalarEq-1})
is equivalent to the equation of motion of a
massive conformally coupled scalar.

The Klein-Gordon equation (\ref{eq:ScalarEq-1}) is separable when effective mass (\ref{eq:effective-mass})
is adopted. By assuming a separation $\varphi=R(r)S(\theta)e^{-i\omega t+im\phi}$,
the angular part of (\ref{eq:ScalarEq-1}) turns out to be
\begin{equation}
\frac{\partial_{\theta}(\sin\theta\Delta_{\theta}\partial_{\theta}S)}{\sin\theta S}-(1+\alpha)^{2}\frac{B^{2}}{\Delta_{\theta}}-2\alpha\cos^{2}\theta=-\lambda,\label{eq:Spheriodal}
\end{equation}
with $\lambda$ the separation constant and $B
\equiv a\omega\sin\theta-\frac{m}{\sin\theta}$.
For Schwarzschild BH, $a=\alpha=0$, the
separation constant $\lambda=l(l-1)$ in which $l$
is the angular momentum number. For more general
BHs, there is no explicit analytical
expression of $\lambda$. For Kerr spacetime,
$\Delta_{\theta}=1$ and the above equation
becomes a spheroidal equation which has been
studied in detail in
\cite{spheroidal-Teukolsky,spheroidal-Seidel,Levear-spheroidal,Separation
constant 2,Kerr-sep-const}. And $\lambda$ can be
expanded as a series of $a\omega$ for small
$a\omega$, whose explicit form can be found in
\cite{spheroidal-Seidel,Kerr-sep-const}. For
Kerr-de Sitter BH, Eq.
(\ref{eq:Spheriodal}) is a generalized spheroidal
equation and the explicit expansion of separation
constant $\lambda$ in small $a\omega$ is
\cite{Separation constant}
\begin{eqnarray}
\lambda & = & l(l+1)+\alpha\left(-l(l+1)+2m^{2}-l^{2}H(l)+(l+1)^{2}H(l+1)\right)\nonumber \\
 &  & +\left[-2m-2m\alpha(1-H(l)+H(l+1))\right]a\omega\label{eq:seperation const}\\
 &  & +[H(l+1)-H(l)+\alpha(H(l+1)-H(l)+2\left((l+1)^{2}H(l+1)-l^{2}H(l)\right)\nonumber \\
 &  & -lH^{2}(l)+(l+1)H^{2}(l+1)-\frac{H(l)H(l+1)}{l(l+1)})](a\omega)^{2}+\mathcal{O}((a\omega)^{3}),\nonumber
\end{eqnarray}
with $H(l) \equiv
\frac{2(l^{2}-m^{2})l}{(2l-1)(2l+1)}$ and $l$
being the angular momentum which takes an integer
or half integer number satisfying $l\geq m$.

The radial part of (\ref{eq:ScalarEq-1}) is
\begin{eqnarray}
\Delta_{r}\frac{d}{dr}(\Delta_{r}\frac{d}{dr}R)+\left[(1+\alpha)^{2}K^{2}-\Delta_{r}\left(\frac{2\alpha}{a^{2}}r^{2}+G(r)+\lambda\right)\right]R & = & 0,\label{eq:radial eq}
\end{eqnarray}
where $K=\omega(a^{2}+r^{2})-am$.

We look for the complex frequencies of the
perturbation $\omega=\omega_R+i\omega_I$. We will
concentrate on looking for unstable modes with
$\omega_I>0$, which signal the instability with
the growing behavior of the perturbations. To
calculate the perturbation frequencies, we need
to impose suitable boundary conditions,
\begin{eqnarray}\label{boundarycondition}
R \rightarrow \Bigg\{\begin{array}{ll}
(r-r_{h})^{-i (\omega-m\Omega_{h})/2\kappa_{h}} &\ \  r\rightarrow r_{h},\\
(r-r_{c})^{i (\omega-m\Omega_{c})/2\kappa_{c}} &\ \  r\rightarrow r_{c}.
\end{array}
\end{eqnarray}
This boundary condition is physically acceptable,
which corresponds to a purely ingoing wave at the
BH event horizon and an outgoing wave at
the cosmological horizon. In asymptotically flat
spacetime, for example the Kerr BH, there
is no cosmological horizon and the outgoing wave
condition is put at the infinity. It was argued
in \cite{Brady:1999wd,Molina:2003dc} that
when the conditions at infinity in asymptotically
flat  BH are altered in de Sitter black
hole, the usual power-law scenario in the late
time tail of the perturbation in asymptotically
flat BH does not necessarily survive in
the de Sitter background. Thus it is of interest
to examine the superradiant stability in the
Kerr-de Sitter configuration and compare with the
result disclosed in the Kerr BH
\cite{Cardoso}.

In general, it is difficult to solve the
perturbation equations analytically. So special
numerical techniques have been developed to
calculate the perturbations of BHs
\cite{QNM-methods}, such as P\"{o}schl-Teller
potential method \cite{P-Teller}, finite
difference method \cite{Finite-diff}, WKB
approximation \cite{WKB}, continued fraction
method \cite{Levear-spheroidal}, direct
integration method (also called shooting method)
\cite{Directly Integ}, etc. In this paper, we
will use the direct integration method to
calculate the frequencies of the perturbations.
The algorithm of this method can be described
briefly as follows: with the boundary conditions
(\ref{boundarycondition}), we can perform two
integrations: (1) solve the radial equation from
the BH event horizon by integration to a
matching point $r=r_{m}$; (2) solve the radial
equation from cosmological horizon to the
matching point. Then by matching the two
integration results, we can derive the
frequencies of the perturbations. The numerical
results are given in the next section.

\subsection{Superradiance condition}

Before doing numerical calculations, let us
briefly derive the superradiant condition for
Kerr-de Sitter BH. Taking the following
variables
\begin{eqnarray}
dr_{\ast}=\frac{r^{2}+a^{2}}{\Delta_{r}}dr & , & \Psi=\sqrt{r^{2}+a^{2}}R,
\end{eqnarray}
the radial equation (\ref{eq:radial eq}) can be written in the form of Schr\"{o}dinger equation
\begin{equation}\label{Schrodingereq}
\frac{d^{2}\Psi}{dr_{\ast}^{2}}+ (\omega^2-V)\Psi=0,
\end{equation}
in which the effective potential reads
\begin{eqnarray}
V & = & \omega^2-\frac{V_{0}}{(r^{2}+a^{2})^{2}}+\frac{\Delta_{r}}{(r^{2}+a^{2})^{3}}\left(r\frac{d\Delta_{r}}{dr}+\frac{a^{2}-2r^{2}}{r^{2}+a^{2}}\Delta_{r}\right),\\
V_{0} & = & (1+\alpha)^{2}K^{2}-\Delta_{r}\left(\frac{2\alpha}{a^{2}}r^{2}+G(r)+\lambda\right).\nonumber
\end{eqnarray}

In a scattering experiment, solution of
(\ref{Schrodingereq}) has the following
asymptotic behavior
\begin{eqnarray}
\Psi \sim \bigg\{\begin{array}{ll}
{\cal T} e^{-i (1+\alpha) (\omega-m \Omega_h) r_\ast} &\ \  {\rm as} \ \ \ r_\ast \rightarrow -\infty (r\rightarrow r_h),\\
e^{-i (1+\alpha) (\omega-m \Omega_c ) r_\ast} + {\cal R} e^{i (1+\alpha) (\omega-m \Omega_c) r_\ast}  &\ \  {\rm as} \ \ \ r_\ast \rightarrow \infty (r\rightarrow r_c).
\end{array}
\end{eqnarray}
The above boundary conditions correspond to an
incident wave of unit amplitude, $e^{-i (1+\alpha)
(\omega-m \Omega_c )}$, plunges in from the
cosmological horizon and gives rise to a
reflected wave of amplitude ${\cal R}$ going back to the
cosmological horizon and a transmitted wave of
amplitude ${\cal T}$ at the BH event horizon.
Considering that the effective potential is real,
so $\Psi^\ast$ (complex conjugate of $\Psi$) is
also a solution of (\ref{Schrodingereq}). The
Wronskian of the two linearly-independent
solutions, $W(\Psi, \Psi^\ast) \equiv \Psi
\frac{d}{dr_\ast} \Psi^\ast- \Psi^\ast
\frac{d}{dr_\ast} \Psi$, is a constant
independent of $r_\ast$. Equating values of the
Wronskian at the BH event horizon and at
the cosmological horizon, we get
\begin{eqnarray}
1-|{\cal R}|^2 = \frac{\omega - m \Omega_h}{\omega - m \Omega_c} |{\cal T}|^2.
\end{eqnarray}
Then we can see that if $(\omega - m
\Omega_h)(\omega - m \Omega_c)<0$, we have
$|{\cal R}|^2>1$. This means that the amplitude of the
reflected wave is larger than that of the
incident wave, and superradiance phenomenon
occurs. So we get the condition to trigger the
superradiance \cite{Khanal1985}
\begin{eqnarray}\label{superradianceregime}
m \Omega_c <\omega<m \Omega_h.
\end{eqnarray}
When there is no cosmological constant, from (24)
and (25), we see that the superradiant condition
goes back to that for the Kerr BH.

\section{Numerical results of the superradiant instability}

We list our numerical results in this section. We
fix the BH mass parameter $M=1$, $n=3$ in
the matter profile and the angular indexes $l=m=1$. There are four free parameters
left to affect the numerical results, namely the
cosmological constant $\Lambda$, the angular
momentum per unit mass $a$, the location of the
matter shell $r_0$ and the coupling between
matter and the scalar field $\beta$. To see the
effect of rotation on the instability, in the
following we will consider three cases with
$a=0.99, 0.7, 0.3$, respectively. And in each
case, we will choose some values of $\Lambda$ to
see the influence of the cosmological constant on
the instability.

\subsection{$a=0.99$}
In this subsection, we fix the angular momentum
per unit mass of the BH and examine the
influences of the other parameters on the
superradiant stability.

In figure 2, we repeat the result of the
superradiant instability in the Kerr BH
background and we find that our numerical results
are in good agreement with that reported in
\cite{Cardoso}. This gives us confidence in
generalizing our numerical calculation  to
examine the stability in the Kerr-de Sitter
backgrounds.

From Figs.3-5, we list the superradiant
instability details for the Kerr-de Sitter black
hole backgrounds with the cosmological constant
$\Lambda=0.0003, 0.03, 0.1$ respectively. The
results are similar to that in the Kerr black
hole case. When the real parts of the
perturbation frequencies satisfy the superradiant
condition (\ref{superradianceregime}), the
imaginary parts of the frequencies have positive
values, indicating the existence of the
superradiant instability. In the figures, since
$\Omega_c$ is too small at the cosmological
horizon, it is not shown.

\begin{figure}[H]
\centering
\includegraphics[width=0.46\textwidth]{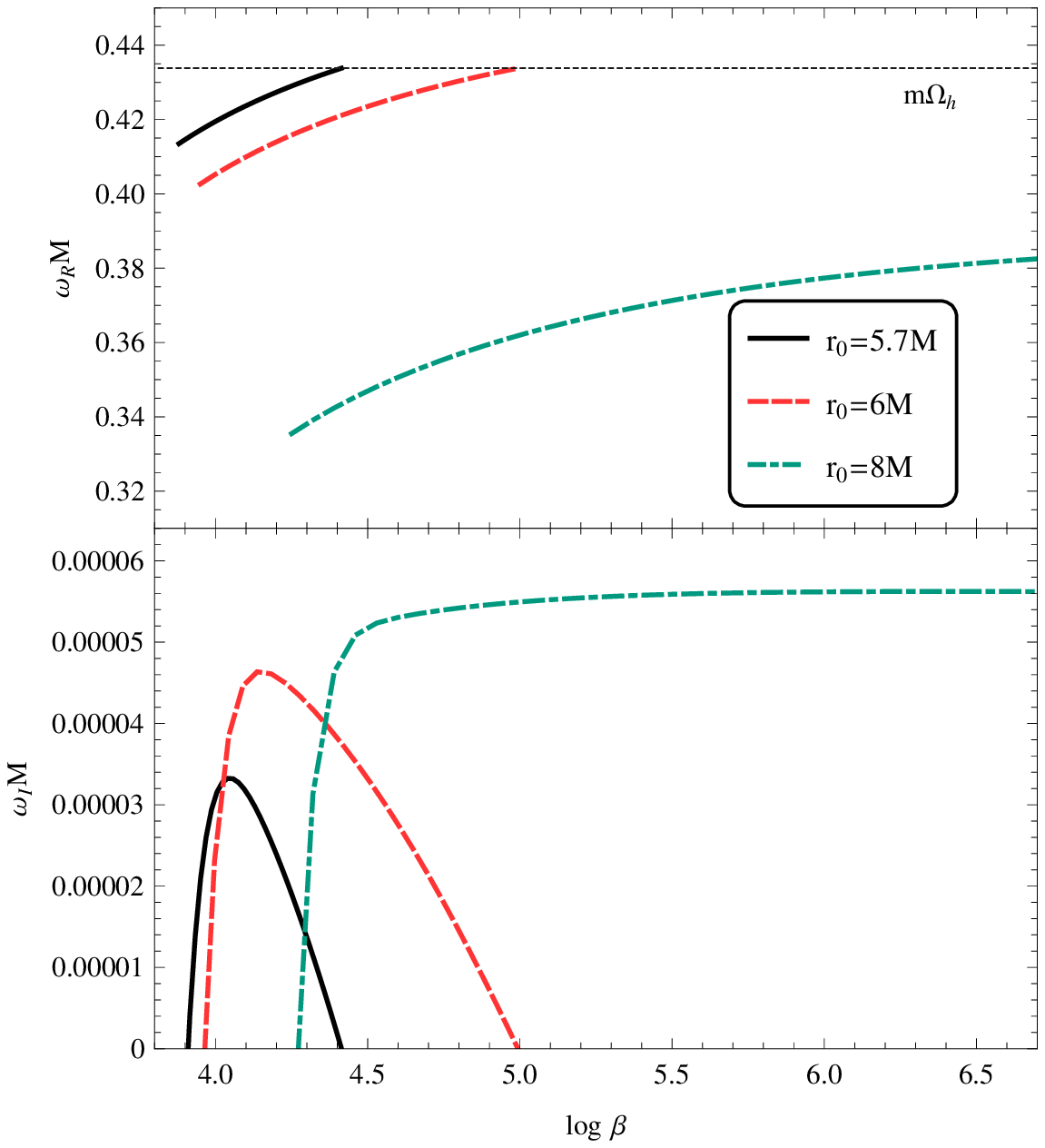}
\includegraphics[width=0.46\textwidth]{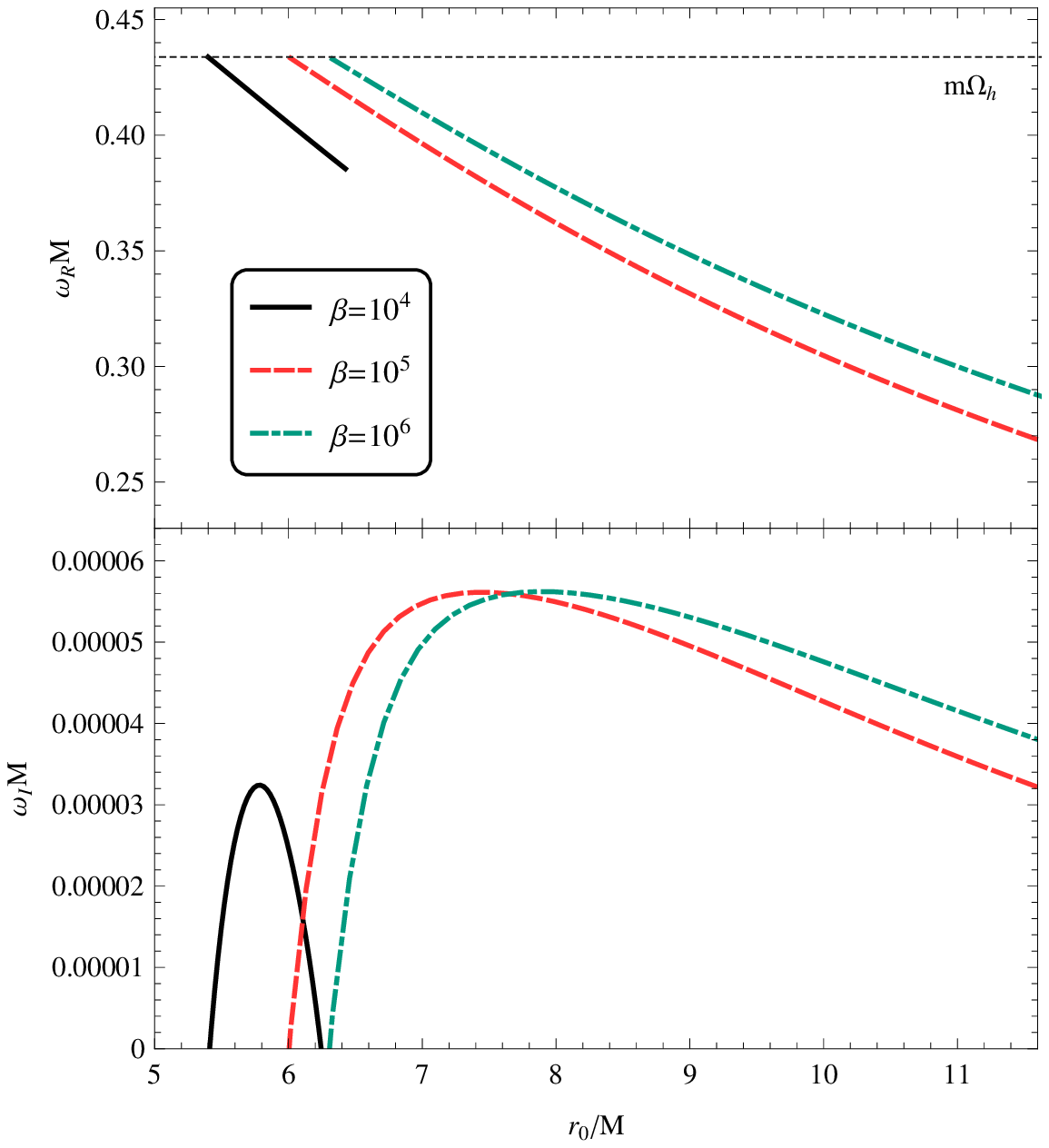}
\caption{Superradiant instability details for
Kerr BH.  The outer BH event
horizon $r_{h}=1.14107$ and the angular velocity
on the outer horizon $\Omega_{h}=0.4338$.}
\end{figure}

\begin{figure}[H]
\centering
\includegraphics[width=0.46\textwidth]{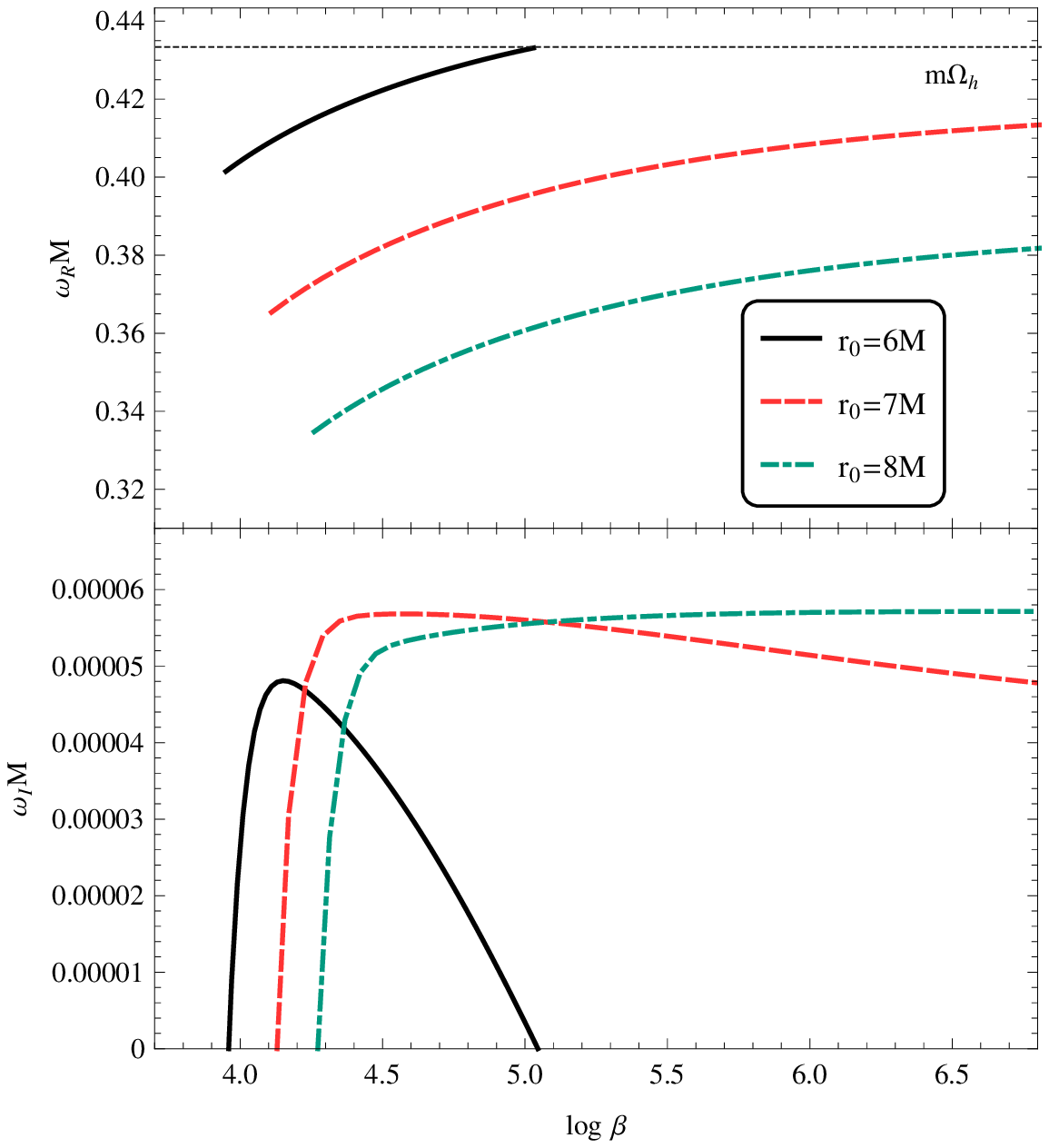}
\includegraphics[width=0.46\textwidth]{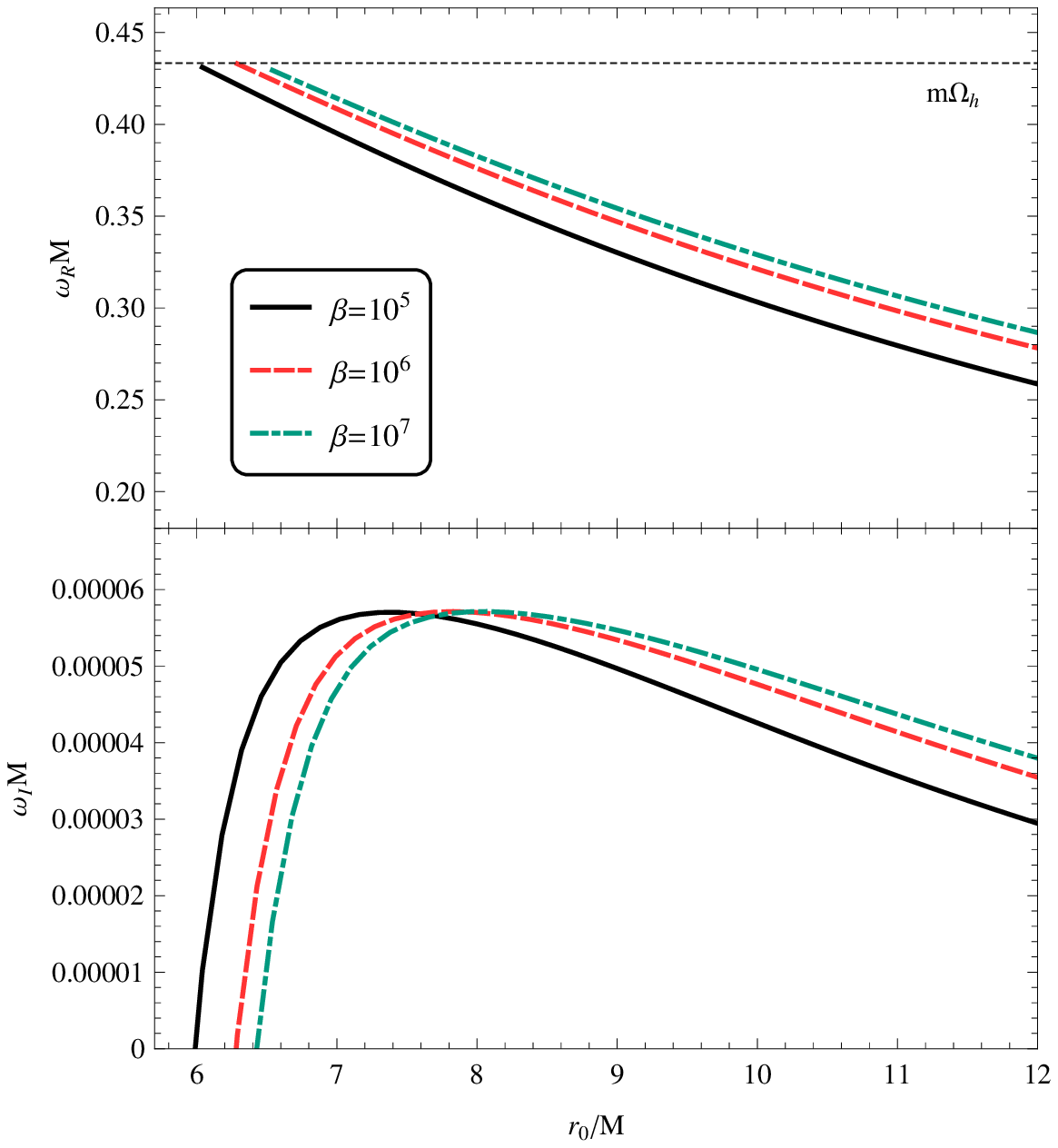}
\caption{Superradiant instability details for
Kerr-de Sitter BH with $\Lambda=0.0003$.
The BH event horizon and cosmological
horizon locate at $r_{h}=1.1421$,
$r_{c}=98.9447$. The angular velocity on the
BH horizon  $\Omega_{h}=0.4333$.}
\end{figure}

\begin{figure}[H]
\centering
\includegraphics[width=0.46\textwidth]{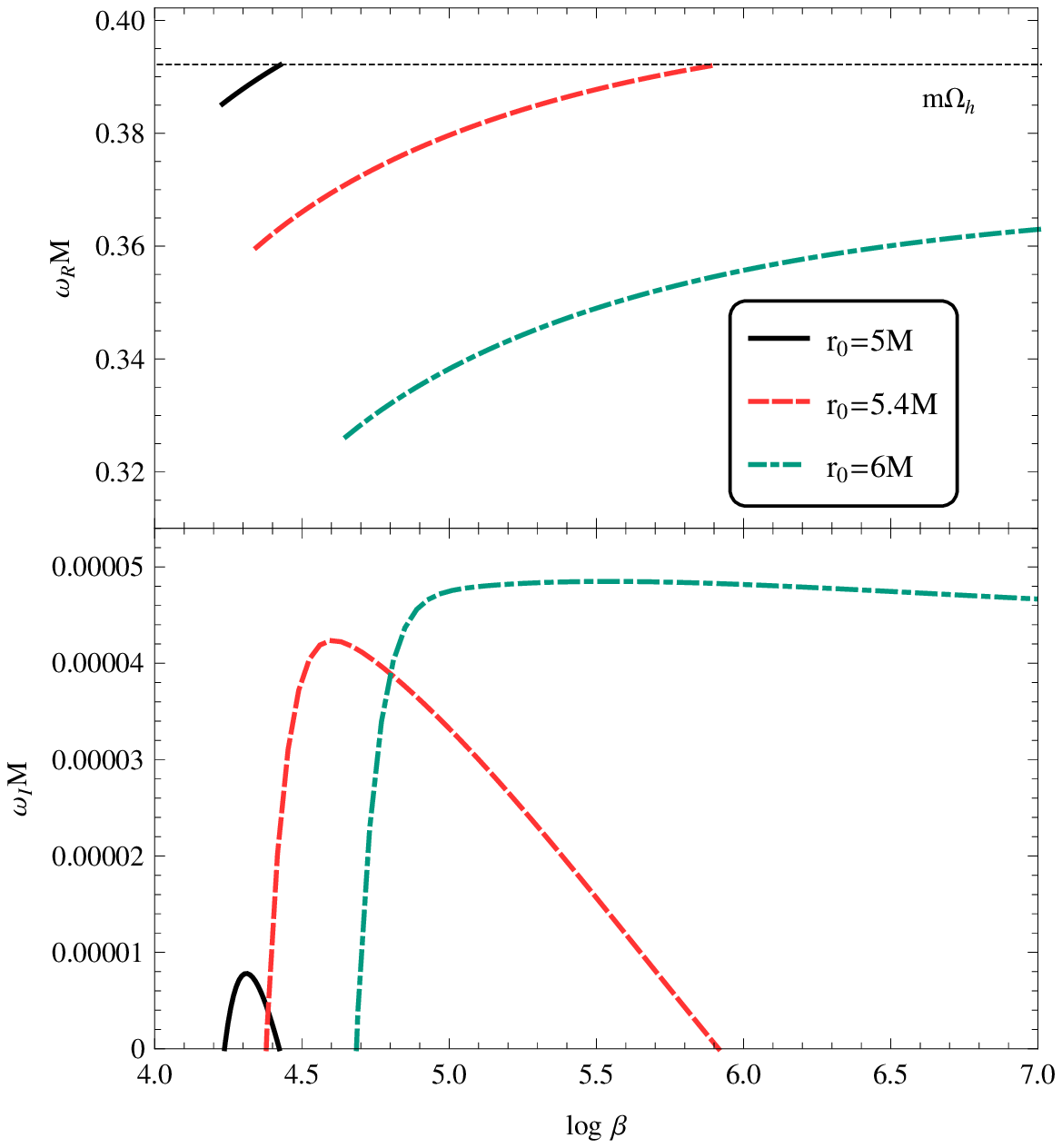}
\includegraphics[width=0.46\textwidth]{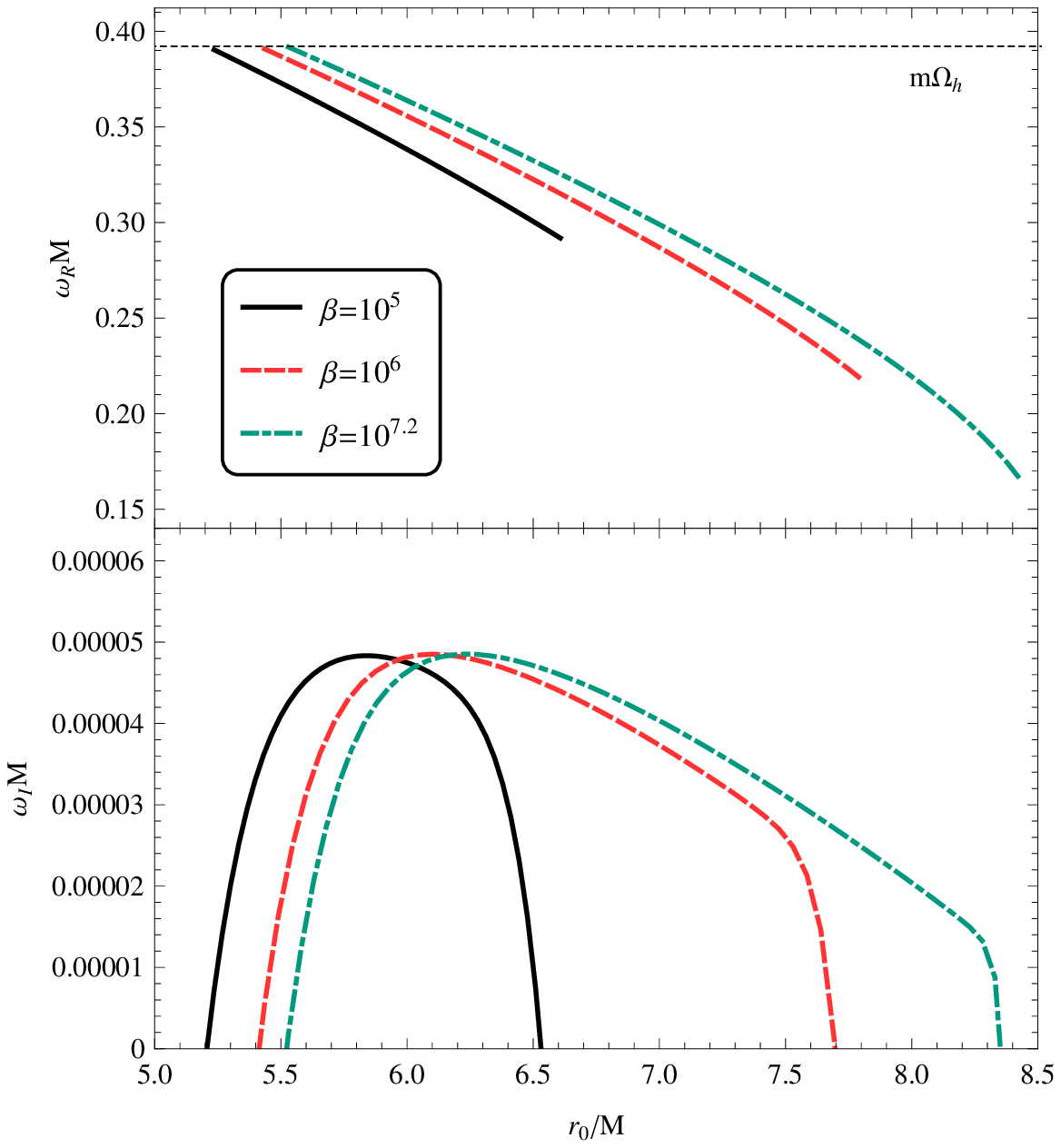}
\caption{Superradiant instability details for
Kerr-de Sitter BH with $\Lambda=0.03$.
The BH event horizon and cosmological
horizon locate at $r_{h}=1.2426$, $r_{c}=8.8077$.
The angular velocity on the BH horizon
$\Omega_{h}=0.3922$.}
\end{figure}

\begin{figure}[H]
\centering
\includegraphics[width=0.46\textwidth]{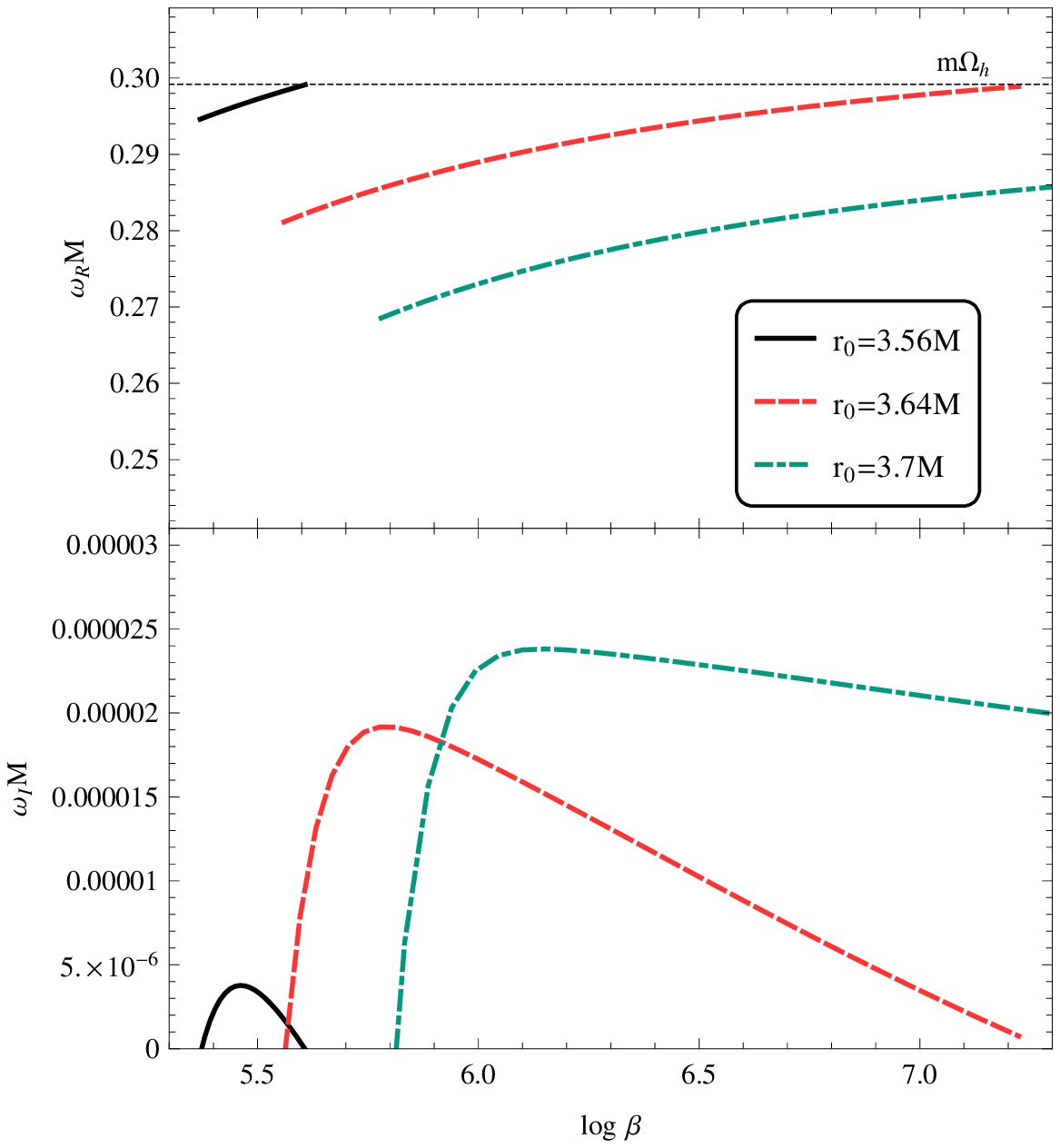}
\includegraphics[width=0.46\textwidth]{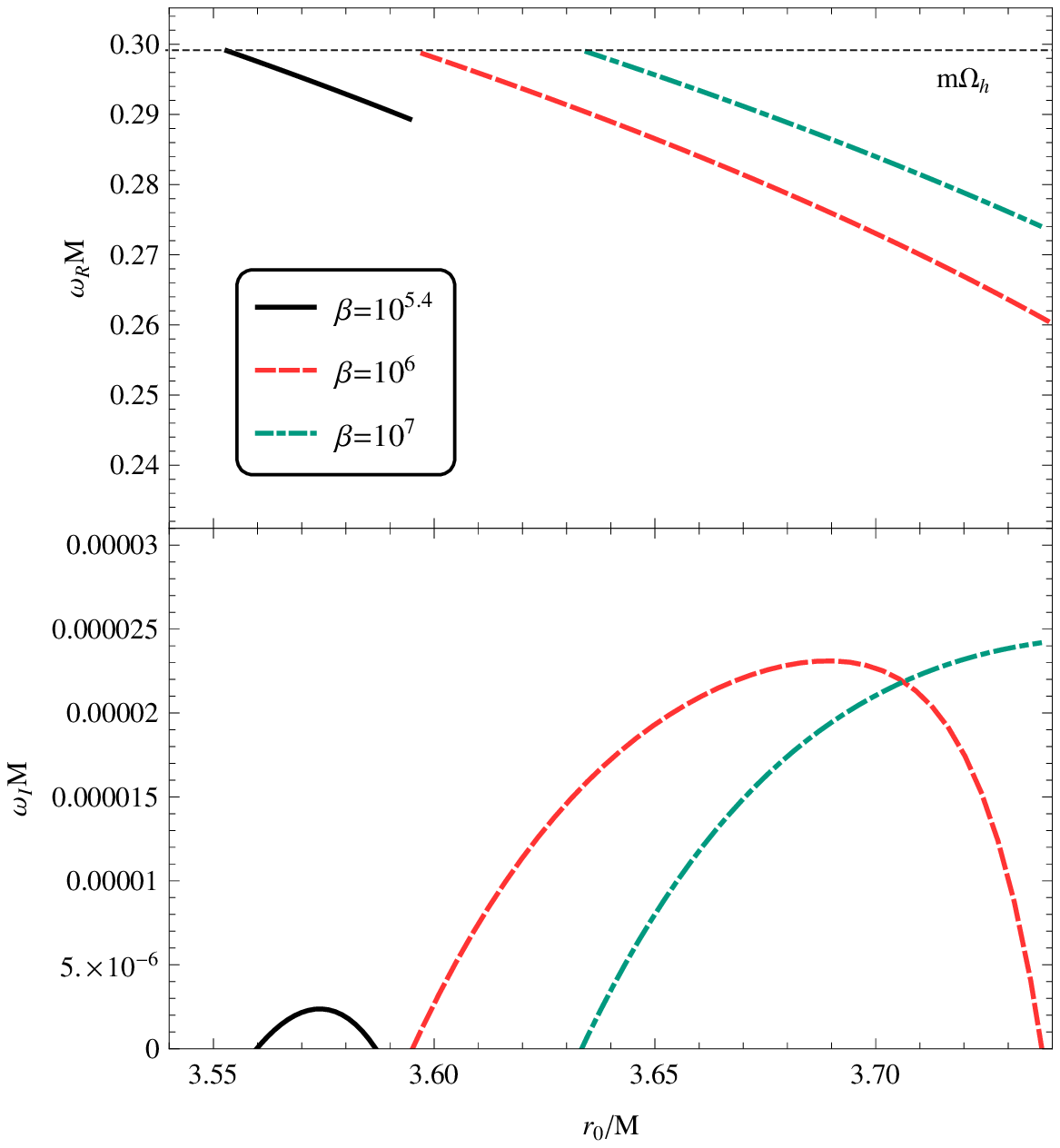}
\caption{Superradiant instability details for
Kerr-de Sitter BH with $\Lambda=0.1$. The
BH event horizon and cosmological horizon
locate at $r_{h}=1.5261$, $r_{c}=3.9722$. The
angular velocity on the BH horizon
$\Omega_{h}=0.2992$.}
\end{figure}

In the Kerr-de Sitter BH backgrounds, we
see that with the increase of the cosmological
constant, the BH event horizon $r_h$
increases, which leads the angular velocity on
the BH event horizon to decrease for the
BH with the same $a$ \cite{QNM of
Kerr-dS,freque-decrease}. The decrease of the
upper bound in the superradiant condition (25)
results in the decrease of the real part of the
perturbation frequencies. This can be seen
clearly from Fig.3-5. In addition, we can read
from Fig.3-5 that the imaginary part of the
frequencies also decreases with the increase of
the cosmological constant in the Kerr-de Sitter
background. Thus the superradiant instability is
quenched with the increase of the cosmological
constant. Physically this can be understood by
looking at the effective potential in Fig.6.
Fixing parameters $a, r_0, \beta$, we see that
with the increase of the cosmological constant in
the Kerr-de Sitter BH background, the
potential well becomes shallower and the barrier
to reflect radiation back becomes lower. This
leads to fewer outgoing waves to be reflected
back and stored in the potential well, which
explains why the superradiant instability is
quenched for large cosmological constant.

Comparing Figs.3-5, we find that to compensate
the influence brought by increasing the
cosmological constant, we need larger $\beta$ or
smaller $r_0$ to trigger the superradiant
instability in the Kerr-de Sitter background.
From (\ref{eq:effective-mass}), we can see that
larger $\beta$ or smaller $r_0$ leads to bigger
effective mass and so that higher effective
potential barrier. Objective pictures of the
effective potentials due to the influences of
$\beta$ and $r_0$ are shown in Fig.7. It is clear
that for fixed $r_0$, the increase of $\beta$
increases the potential barrier height which is
able to reflect more radiations back. Fixing
$\beta$, the decrease of $r_0$ has the same
effect. Besides if we look closely at the
potentials in Fig.7 near the BH event
horizon, the first potential barrier is higher so
that the developed potential well is deeper for
bigger $\beta$ as well as for smaller $r_0$.
Thus more outgoing wave can be reflected back by
the higher potential wall and accumulated in the
deeper potential well. This accounts for the
phenomenon that with the increase of the
cosmological constant we need to count on bigger $\beta$ and
smaller $r_0$ to spark off the superradiant
instability.

From Figs.3-5 we also observe some similar
properties to those of the Kerr BH
reported in \cite{Cardoso}. When the matter shell
is too close to the Kerr-de Sitter BH
with very small $r_0$, there is no superradiant
instability, since the real part of the
frequencies scales as $1/r_0$ and too small $r_0$
will make the frequency violate the superradiant
condition. If $r_0$ is not small enough, the
superradiant instability can appear, but this
instability will be quenched as soon as the
superradiant condition is saturated. The value of
$r_0$ where the instability starts to appear decreases
with the increase of the cosmological constant
adopted. When $r_0$ is big, the positive
imaginary part of the frequencies can last for
sufficiently large $\beta$. This is because that
once $\beta$ is big enough to provide high
enough potential barrier to bounce back the
outgoing wave, further increasing $\beta$ to raise
the height of the potential wall will not
contribute more to the instability. We find that
the above mentioned properties keep for all
values of the cosmological constant in the
Kerr-de Sitter BH background.

\begin{figure}[H]
\centering
\includegraphics[width=0.4\textwidth]{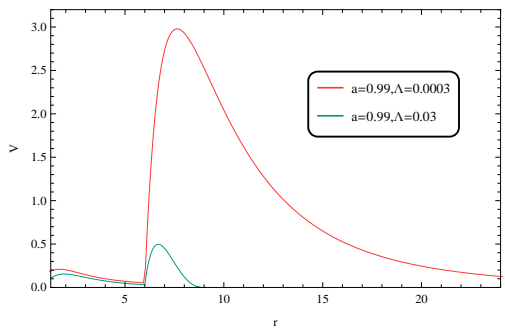}
\caption{Effective potential behavior for the
change of the cosmological constant. We fixed
$a=0.09, \beta=10^{4.81}, r_0=6M$. }
\end{figure}

\begin{figure}[H]
\centering
\includegraphics[width=0.4\textwidth]{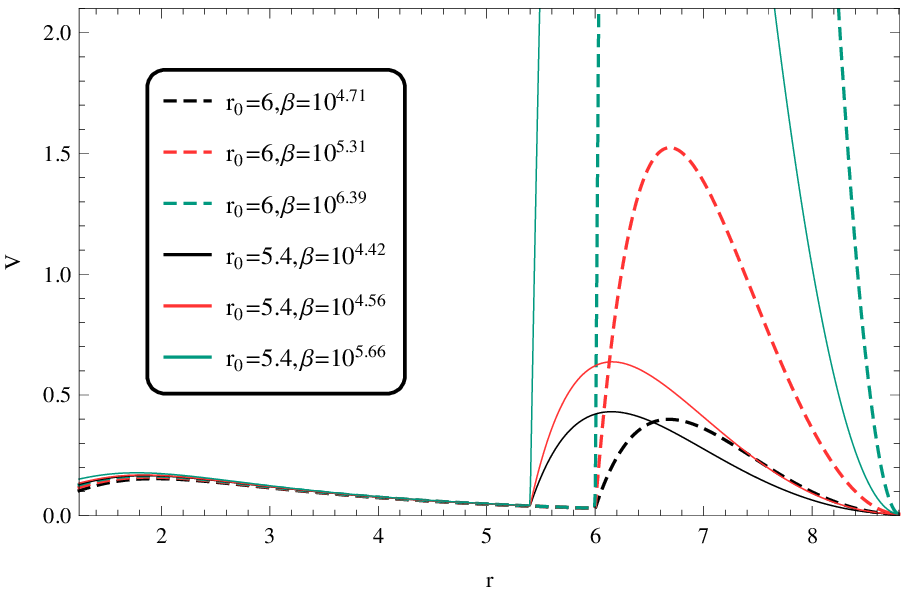}
\includegraphics[width=0.4\textwidth]{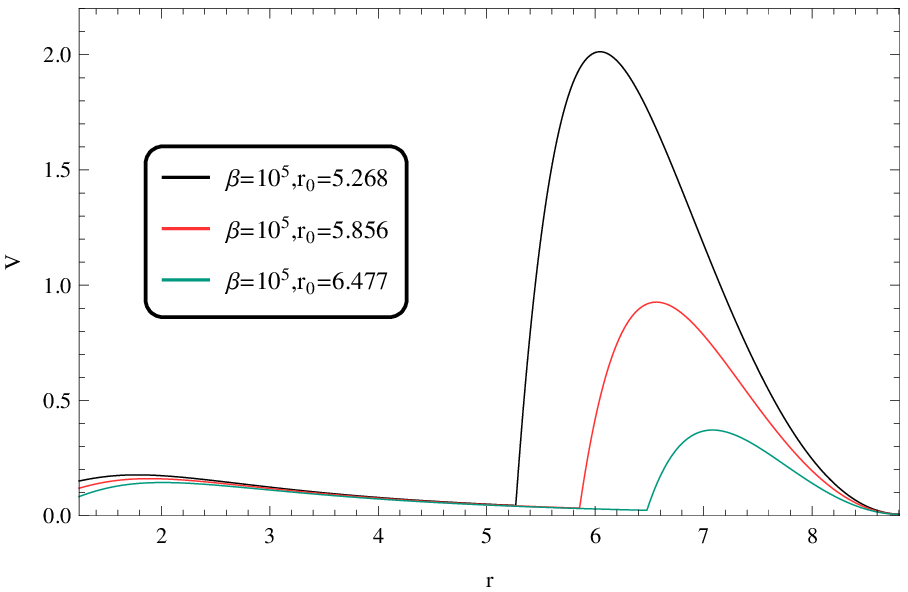}
\caption{{\em (Left)} Effective potential
behavior for the change of $\beta$. We fixed
$a=0.09, \Lambda=0.03$. The solid lines are for
$r_0=5.4$ and dashed lines are for $r_0=6$.  {\em
(Right)} Effective potential behavior for the
change of $r_0$. We fixed $a=0.99, \Lambda=0.03,
\beta=10^{5}$. }
\end{figure}

\subsection{$a=0.7$ and $a=0.3$}

In this subsection  we will fix $\Lambda= 0.03$
and show the results of superradiant instability
for $a=0.7$ and $a=0.3$. Combining Figs.8,9 and
Fig.4, we can show the influence of the angular
momentum per unit mass on the superradiant
instability for the Kerr-de Sitter BH
background.

\begin{figure}[H]
\centering
\includegraphics[width=0.46\textwidth]{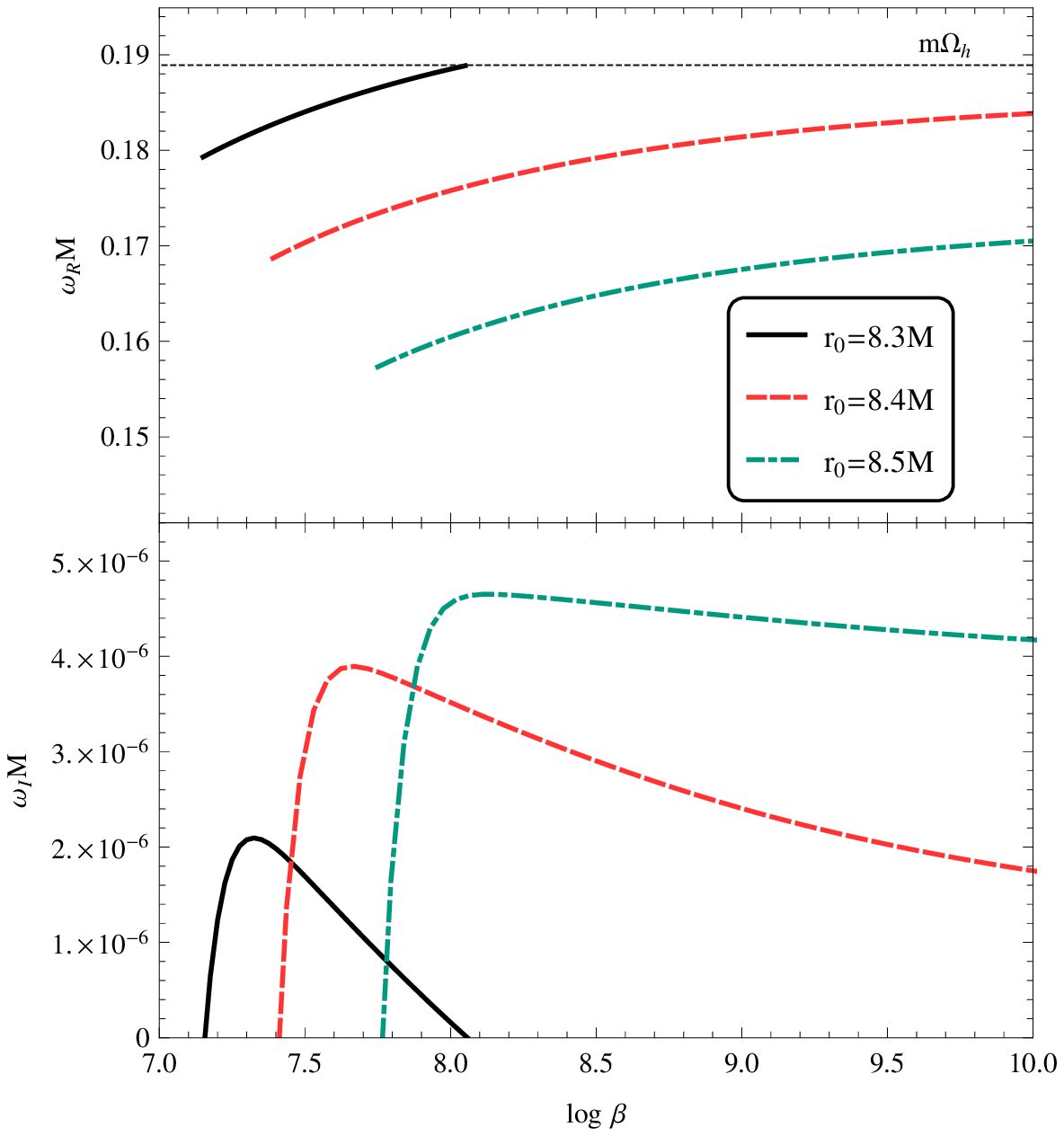}
\includegraphics[width=0.46\textwidth]{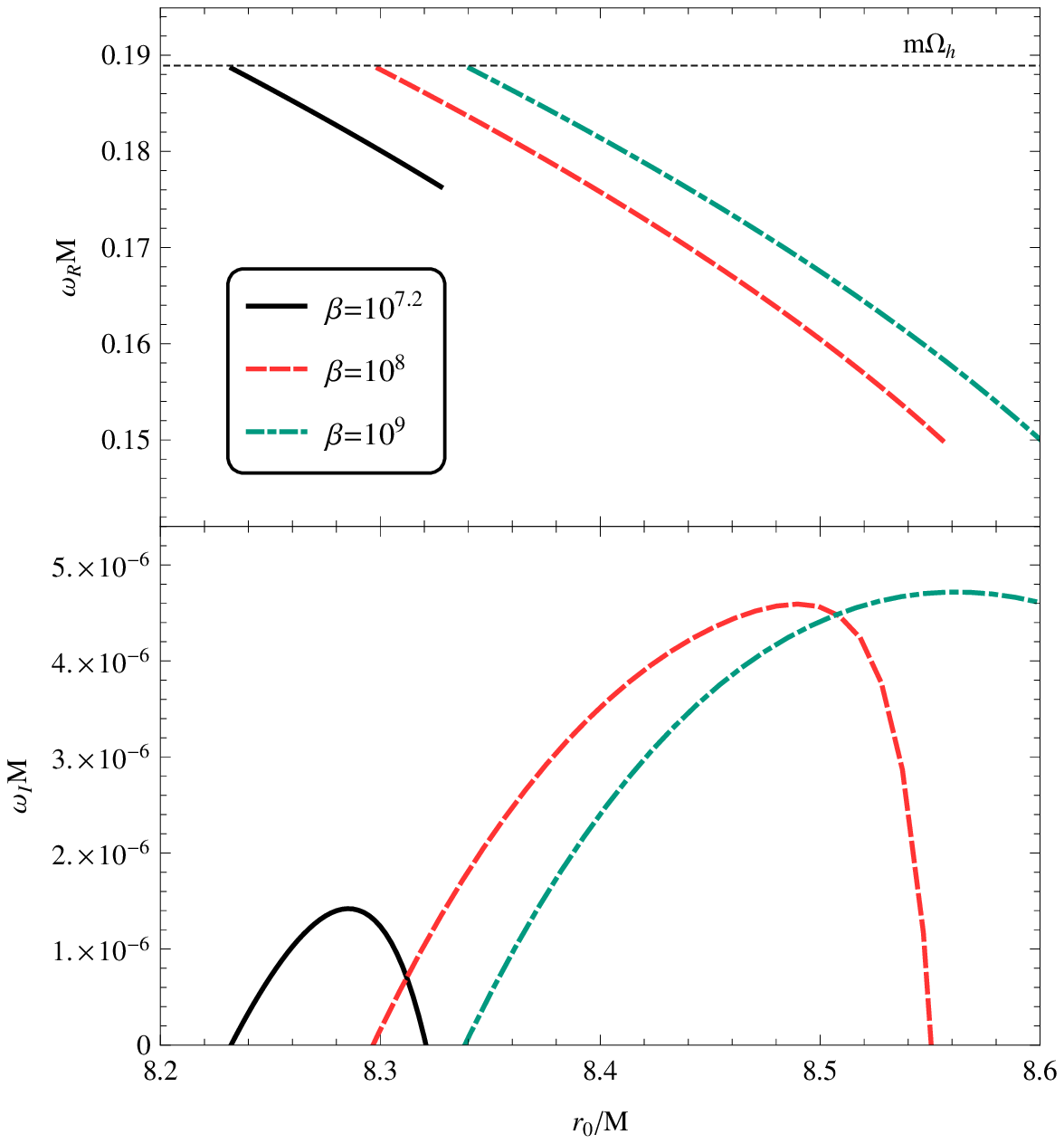}
\caption{Superradiant instability details for
Kerr-de Sitter BH with $a=0.7,
\Lambda=0.03$. The BH event horizon and
cosmological horizon locate at $r_{h}=1.7932$,
$r_{c}=8.7984$. The angular velocity on the black
hole horizon $\Omega_{h}=0.1889$.}
\end{figure}

\begin{figure}[H]
\centering
\includegraphics[width=0.46\textwidth]{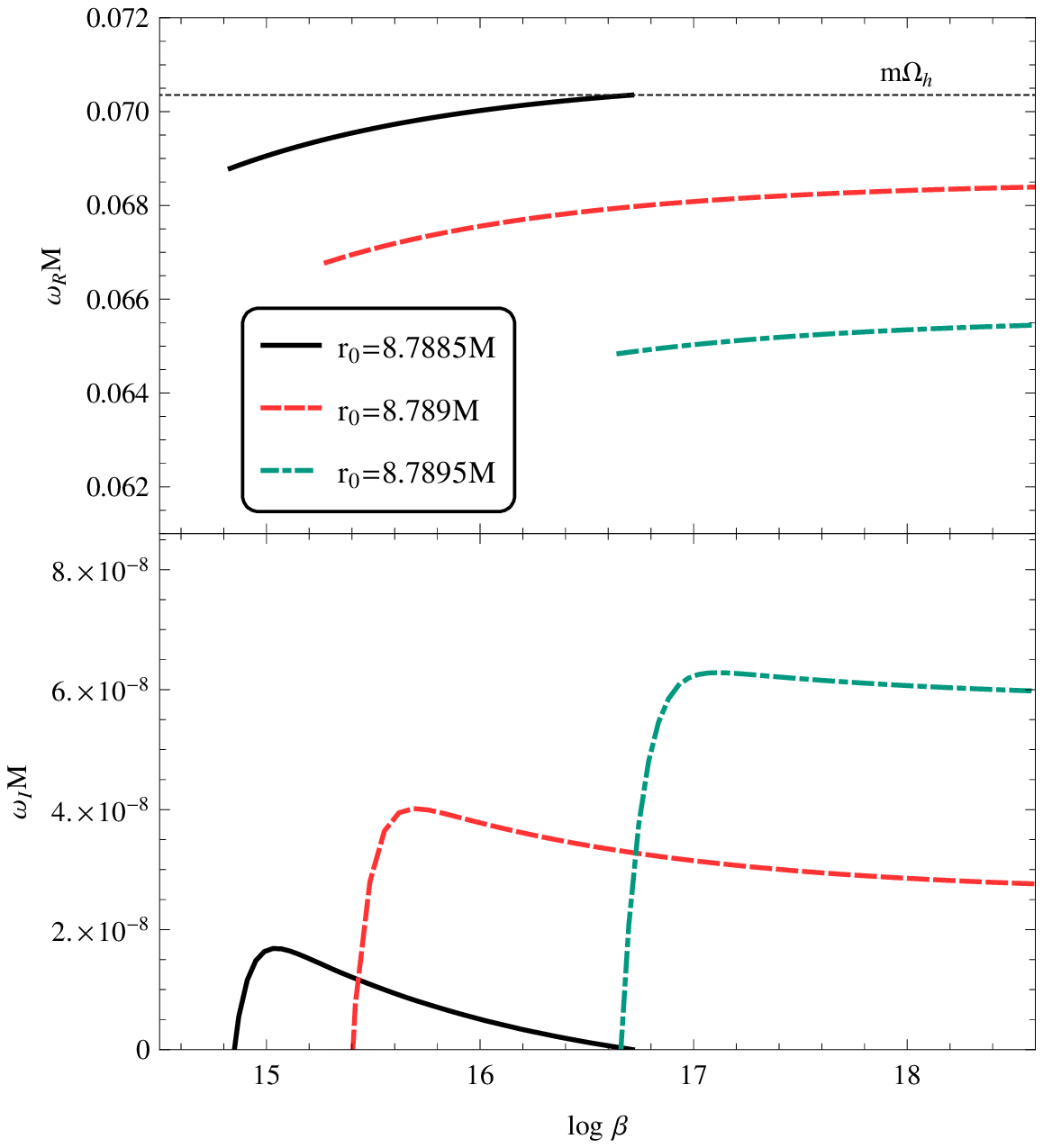}
\includegraphics[width=0.46\textwidth]{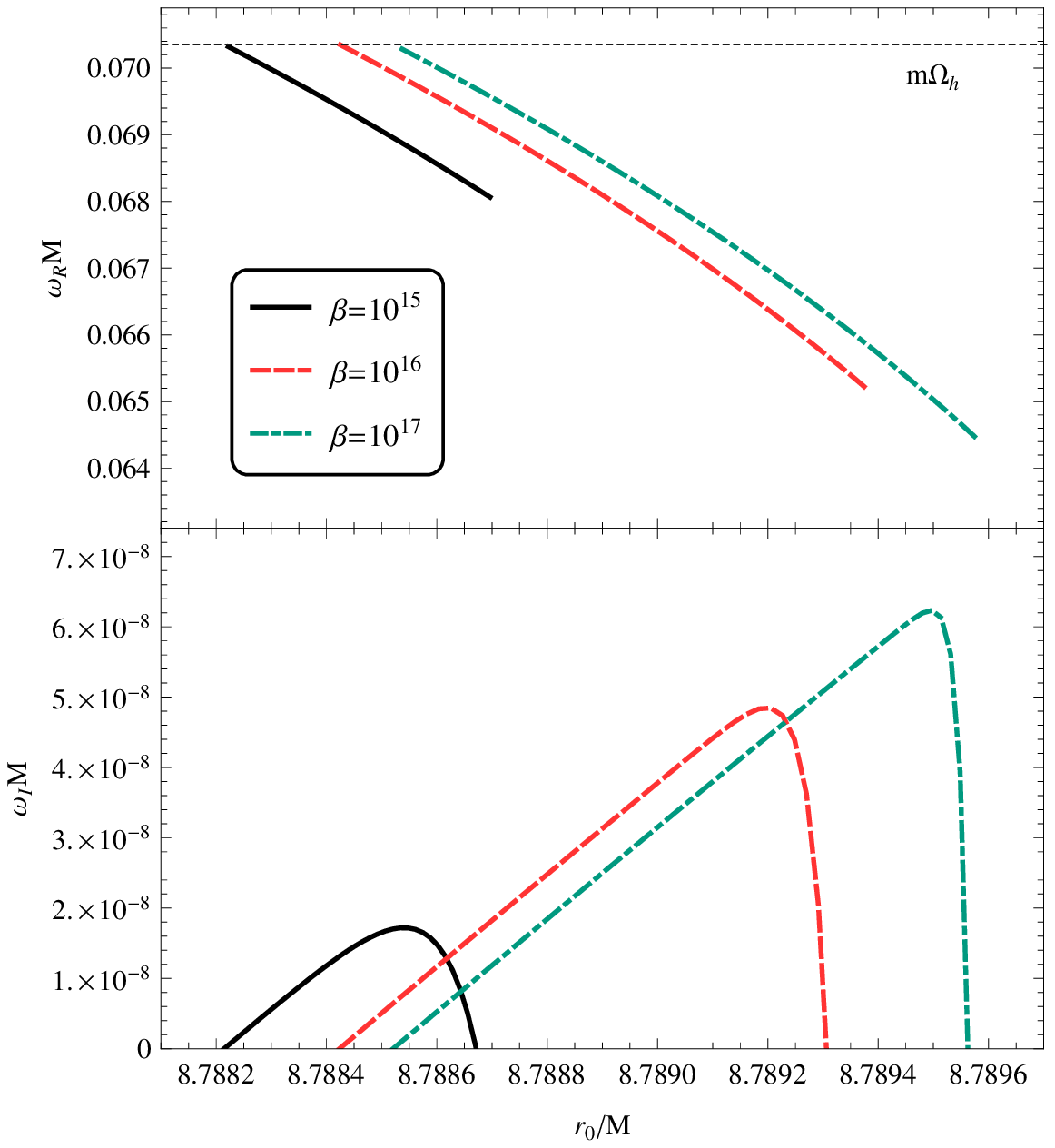}
\caption{Superradiant instability details for
Kerr-de Sitter BH with $a=0.3,
\Lambda=0.03$. The BH event horizon and
cosmological horizon locate at $r_{h}=2.0431$,
$r_{c}=8.7906$. The angular velocity on the black
hole horizon $\Omega_{h}=0.0704$. }
\end{figure}

From Figs.8,9 and combining with Fig.4 for
$a=0.99$, we can see that when $a$ decreases, the
upper limit for the superradiant condition
decreases which forces the upper value of the
real part of the frequency to decrease.
Considering that the real part of the frequency
scales with $1/r_0$, we learn that for the
Kerr-de Sitter BH with the same
cosmological constant but slower rotation, the
superradiant instability can occur when the
matter shell is put a bit further away with
bigger $r_0$. Furthermore we observe that with
the decrease of $a$, the imaginary part of the
frequency falls closer to zero, which indicates
that the instability becomes milder. This is
understandable because when the BH
rotates slower, it will be harder to extract the
rotational energy through superradiance. Fixing
$\Lambda, r_0, \beta$, we show the potential
behavior with the change of $a$ in Fig. 10. It is
clearly shown that with the decrease of $a$, the
potential barrier to reflect the outgoing wave
becomes lower and the potential well becomes
shallower. This can be used to account for the
observation that lower $a$ makes the superradiant
instability milder. We further exhibit the
property of the potential when the imaginary part
of the frequency reaches the peak for small
$\beta$ in Figs.4,8,9 in the left panel of
Fig.12. For the limiting $(r_0, \beta)$ to
trigger the strongest superradiant instability,
we see that for the lower $a$ we have shallower
potential barrier to store the reflected
perturbation. This again explains the reason that
the superradiant instability is weaker when $a$
is smaller.

\begin{figure}[H]
\centering
\includegraphics[width=0.4\textwidth]{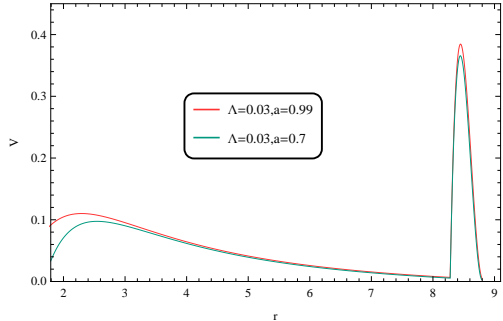}
\caption{Effective potential behavior with the
change of $a$, where we fix $\Lambda=0.03,
\beta=10^{7.2}, r_0=8.28$.}
\end{figure}

\subsection{The minimum $a$ to accommodate the superradiant instability}

In the last subsection, we observe that lowering
$a$ will make superrandiant instability milder.
We can see this from Fig.9 where $a=0.3$ and
$\Lambda=0.03$. In that case, we need large
enough $\beta$ to trigger instability. The
physical reason is that it is harder to extract
the rotational energy from the Kerr-de Sitter
BH background when $a$ is small.

Now the question comes, whether there is the
minimum value of $a$ to allow the superradiant
instability? We have calculated the result, for
$\Lambda=0.1$, $a_{min}\sim0.54$. We calculate the corresponding unstable modes for this critical case,
and the results are shown in Fig.12.
Comparing to Fig.5 with $a=0.99$,
we can see that $\beta$ is needed to
be very large to spark off superradiant
instability and the instability is very mild with very small $\omega_I$. Below this
$a_{min}$, no matter how big is the coupling
between the matter and the scalar, the
superradiant instability will not happen.
For the
Kerr BH background, we have $a_{min}\sim 0.18$. We see that $a_{min}$
increases with the increase of the cosmological
constant $\Lambda$. This is reasonable because
the increase of $\Lambda$ makes the superradiant
instability harder to occur as we discussed
above. From the right panel of Fig.11, we
illustrate the potential behavior of the minimum
$a$ for $\Lambda=0.1$ to spark off superradiant
instability. We see that at the minimum $a$, the
potential is very flattened and the potential
well to store the reflected outgoing perturbation
is very shallow. Although the potential wall is
very high to bounce back the outgoing
perturbation, the perturbation will be harder to
accumulate near the BH to cause the
superradiant instability. This accounts for the very mild instability at this critical case. Below this $a_{min}$,
all the reflected outgoing perturbation will fall
inside the BH without any obstacle.

\begin{figure}[H]
\centering
\includegraphics[width=0.46\textwidth]{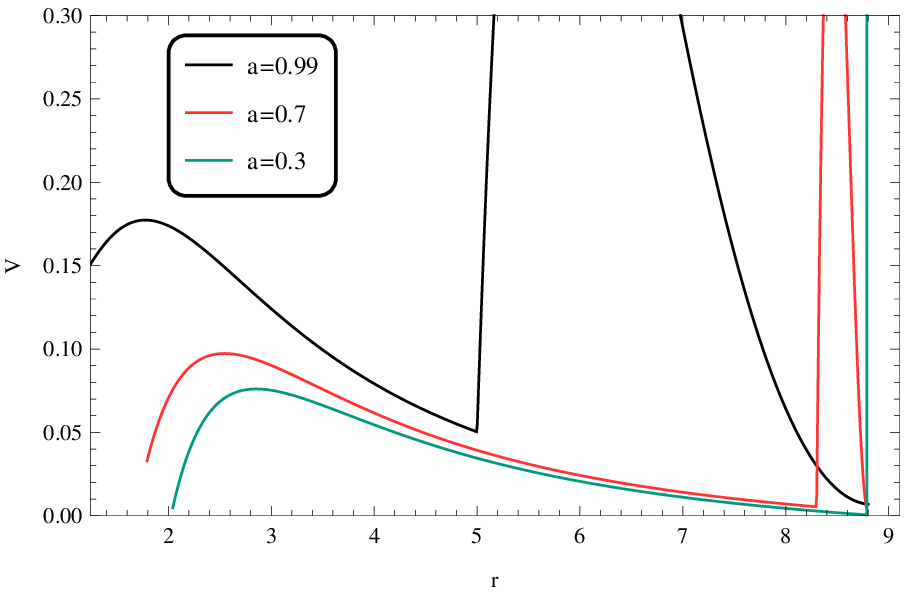}
\includegraphics[width=0.46\textwidth]{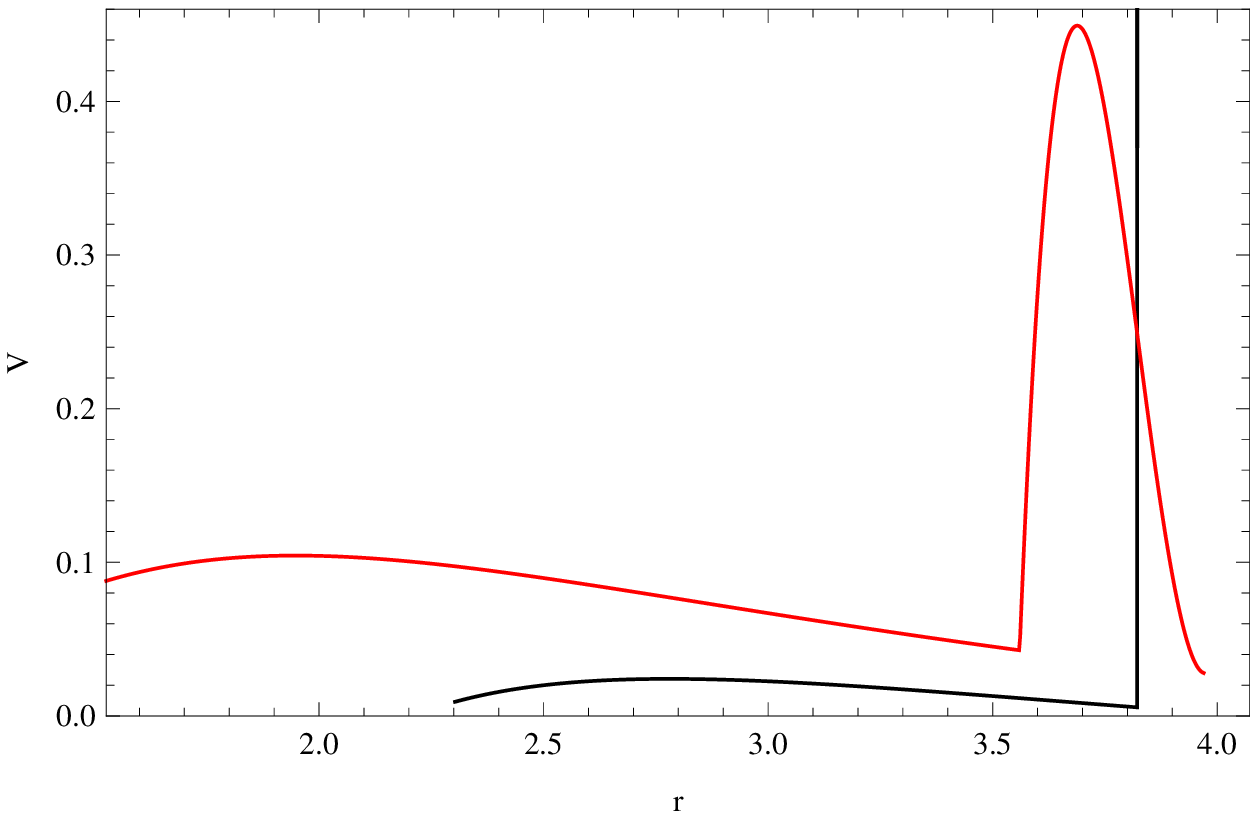}
\caption{{\em (Left)}Effective potential
behaviors with the change of $a$. We fix
$\Lambda=0.03$ and choose $r_0, \beta$ values to
have the smallest peak of $\omega_I$ in
Figs.4,8,9.  For $a=0.99$, we choose $(r_0,
\beta)=(5, 10^{4.31})$; for $a=0.7$, we select
$(r_0, \beta)=(8.3, 10^{7.325})$; and for
$a=0.3$, we adopt $(r_0, \beta)=(8.7885,
10^{15.03})$. {\em (Right)}Effective potential
behavior for the minimum $a$ to accommodate
superradiant instability when $\Lambda=0.1$ (the black one). For
comparison we also plot the line when $a=0.99$
and $\Lambda=0.1$ (the red one). We choose $(r_0,
\beta)=(3.56, 10^{5.46})$ for $a=0.99$ and $(r_0,
\beta)=(3.82156, 10^{14.3879})$ for $a=0.54$,
respectively. }
\end{figure}

\begin{figure}[H]
\centering
\includegraphics[width=0.46\textwidth]{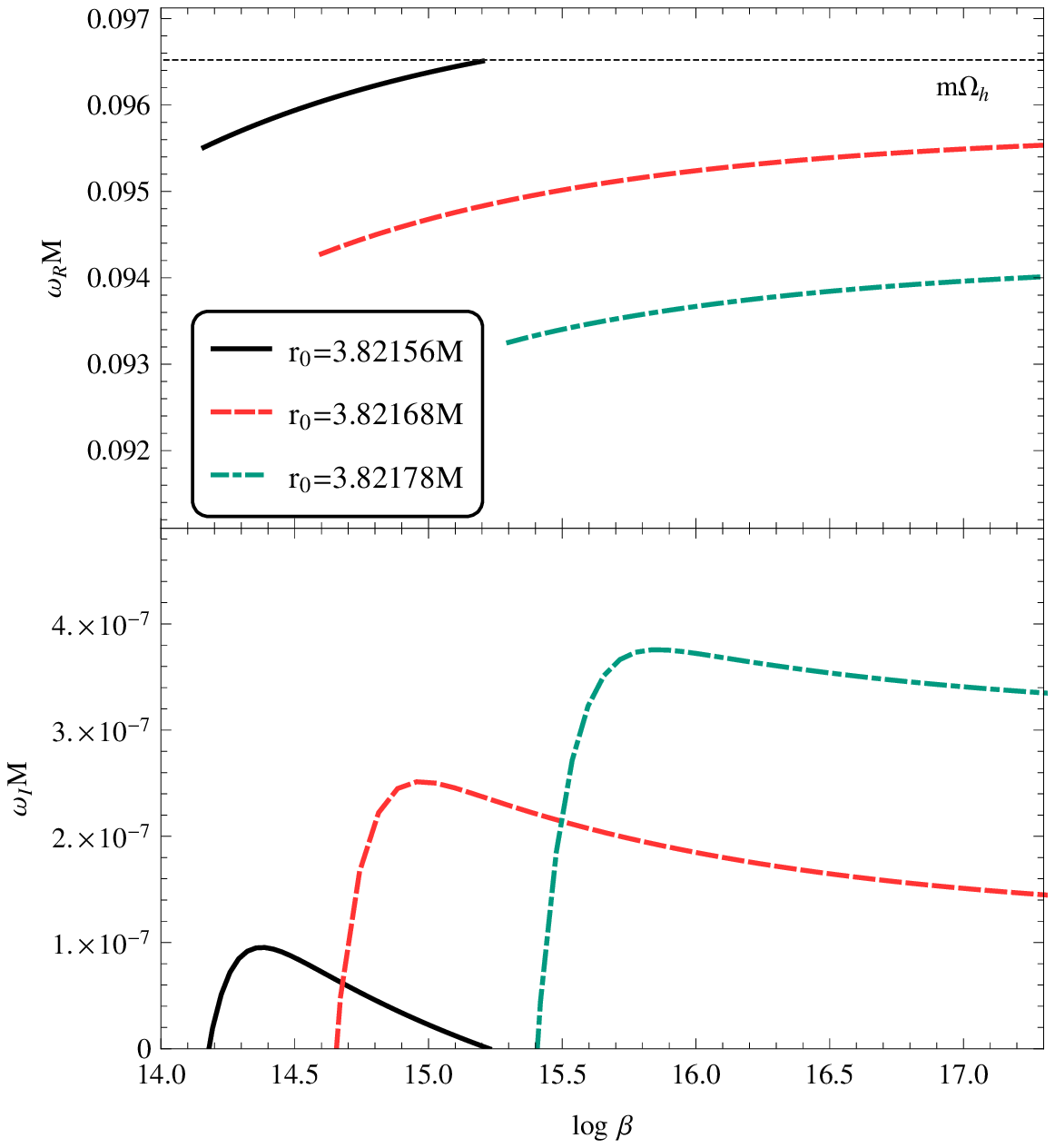}
\includegraphics[width=0.46\textwidth]{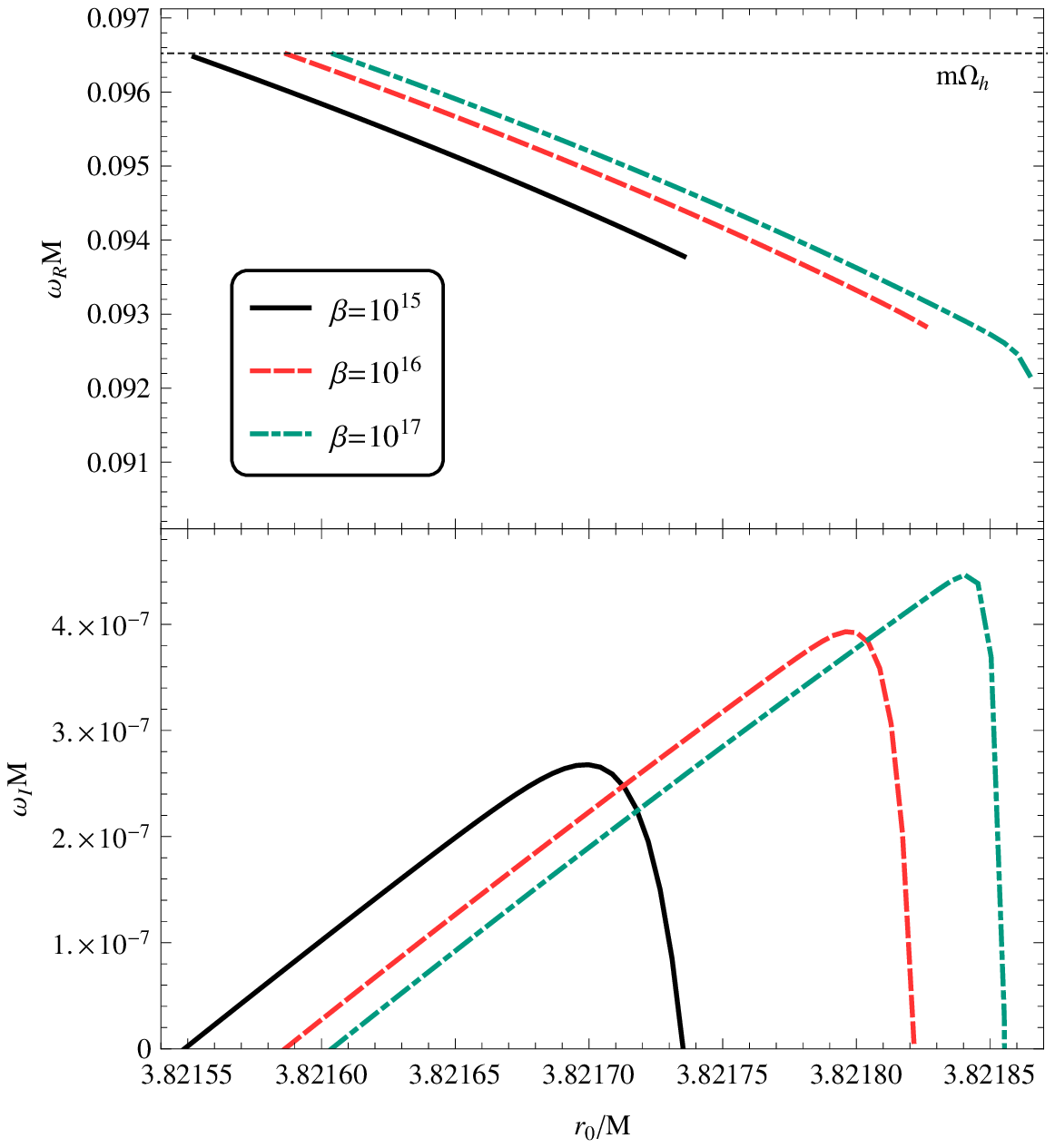}
\caption{\label{fig:alpha4n3}Superradiant
instability details for Kerr-de Sitter BH
with $a=0.54, \Lambda=0.1$. The BH event
horizon and cosmological horizon locate at
$r_{h}=2.30281$, $r_{c}=3.82226$. The angular
velocity on the BH horizon
$\Omega_{h}=0.0965227$.}
\end{figure}

\section{Spontaneous Scalarization}

In \cite{Cardoso}, the authors uncovered a new
instability caused by the distribution of matter
around BHs. When the matter configuration
is dense enough, the BH is forced to
develop scalar hair. This is called the
spontaneous scalarization. In their argument,
they considered that the matter distribution is
spherically symmetric and its backreaction on the
geometry is negligible. In this probe limit the
background metric is a Schwarzschild BH.
They have constructed nonlinear, hairy solutions
of ST theories with a Schwarzschild
BH at the center.

In this section, we will generalize their
argument to the spherically symmetric de Sitter
BHs.  We focus on the four-dimensional
Schwarzschild-de Sitter BH at the center
sounded by spherical shell of scalar field.
Adopting the decomposition of the scalar field
$\varphi(t,r,\theta,\phi)=\sum_{lm}e^{-i\omega
t}Y_{lm}(\theta,\phi)\frac{\Psi_{lm}(r)}{r}$, the
radial wave equation of the scalar field obeys
\begin{eqnarray}
\frac{d^{2}\Psi_{lm}(r)}{dr_\ast^{2}}+[\omega^{2}-V(r)]\Psi_{lm}(r) & = & 0,
\end{eqnarray}
in which
\begin{eqnarray}
V(r) & = & f\left(\frac{l(l+1)}{r^{2}}+\frac{1}{r}\frac{df}{dr}+\mu_{s}^{2}(r)\right),\label{eq:EffPoten}\\
f & = & 1-\frac{2M}{r}-\frac{\Lambda}{3}r^{2}.\nonumber
\end{eqnarray}
Here we have taken the tortoise coordinate
defined as $dr/dr_{\ast}=f$. $\Psi_{lm}(r)$ is
required to satisfy boundary conditions, namely
the outgoing wave at the cosmological horizon and
ingoing wave at the BH event horizon. The
frequency $\omega$ is complex, $\omega=\omega_R+i\omega_I$. The unstable modes
correspond to $\omega_I>0$.

According to the well known result in quantum
mechanics \cite{Sufficient}, the sufficient
condition for this potential to lead to an
instability is
$\int_{-\infty}^{\infty}V(r)dr_{*}=\int_{r_{h}}^{r_{c}}
\frac{V(r)}{f}dr<0.$ Combining with
(\ref{eq:EffPoten}), this condition becomes
\begin{eqnarray}
\int_{r_{h}}^{r_{c}}\mu_{s}^{2}(r)dr & < & -l(l+1)(\frac{1}{r_{c}}-\frac{1}{r_{h}})-M\left(\frac{1}{r_{c}^{2}}-\frac{1}{r_{h}^{2}}\right)-\frac{2\Lambda}{3}(r_{c}-r_{h}).\label{eq:instability-condition}
\end{eqnarray}
It is obvious that this sufficient condition can
be satisfied if the effective mass square is
sufficiently negative. It is straightforward to
prove that for higher dimensional and charged de
Sitter BHs, the instability condition can
also be fulfilled if the effective mass square is
negative enough.

To understand the instability due to the
spontaneous scalarization in the presence of
matter, we focus on the final state of
spherically symmetric configurations described by
\begin{eqnarray}\label{finalstate}
ds^{2} & = &
-h(r)dt^{2}+\frac{1}{f(r)}dr^{2}+r^{2}d\Omega^{2}.
\end{eqnarray}
The scalar field $\Phi$ can be integrated from
the Klein-Gordon equation, obeying
$\partial_{r}\Phi=\frac{Q}{r^{2}\sqrt{fh}}$,
where $Q$ is the scalar charge (the integral
constant). The BH event horizon
($f(r_{h})=0$) is surrounded by the matter shell,
regularity at the BH event horizon requires $Q=0$ and
$\Phi=\Phi_-=const$ inside the matter shell. For the same reason,
at the cosmological horizon out of the matter
shell, $f(r_{c})=0$, which also demands
$\Phi=\Phi_+=const$ there. So the scalar field can only
be constants both inside and outside the matter
shell. However, these two constants, $\Phi_-$ and $\Phi_+$, can be different due to the existence of the matter shell.
That is, there is a finite jump of the scalar field over the matter shell, which leads to a nontrivial configuration. This configuration is very different from that discussed in \cite{Cardoso} without cosmological horizon but only BH event
horizon in the final spherical configuration, which does not have regularity condition out of the matter shell so that the
scalar field $\Phi$ is not limited to be a
constant out of the shell and can
develop a coordinate-dependent profile \cite{Cardoso}.

In our case, due to the constant profile of the scalar field inside and outside the matter shell, the final metric on both sides will take Schwarzschild-de Sitter forms but with different parameters $(M, \Lambda)$. The cosmological constants $\Lambda_\pm$ on both sides are related to the scalar field by the relation $\Lambda_\pm = V(\Phi_\pm)$. If the matter shell is made of perfect fluid, the surface stress-energy tensor takes the form
\begin{eqnarray}
S_{ab}^E = \sigma u_a u_b + P (\gamma_{ab} +u_a u_b),
\end{eqnarray}
where $\sigma, P$ denote density and pressure respectively. $u_a$ is the on-shell four velocity and $\gamma_{ab}$ is the induced metric on the shell. According to the Israel-Darmois junction condition, the jump of the metric over the matter shell can be expressed in terms of the shell composition. In particular, for a static shell at $r=R$, we have \cite{Cardoso,Israel:1966rt}
\begin{eqnarray}
\sigma &=& -\frac{1}{4\pi R} \left(\sqrt{f_+} - \sqrt{f_-}\right)\\
P &=& \frac{1}{8\pi R} \left(-4\pi R\sigma+\sqrt{f_+} \frac{R f_+'}{2 f_+} -\sqrt{f_-} \frac{R f_-'}{2 f_-}\right).
\end{eqnarray}

In summary, in spherical de Sitter BH background, spontaneous scalarization can also happen but in a special style. We should note that the jump of the scalar field over the matter shell discussed above is due to the simplification of the thin shell. In a more realistic configuration, matter will have non-vanishing support as treated in \cite{hair}.

\section{Summary and discussion}

In this paper, we have studied the stability of
Kerr-de Sitter BH surrounded by matter
shell in ST theories. We have found
that similar to the asymptotically flat Kerr
BH \cite{Cardoso}, there exists
superradiant instability in the Kerr-de Sitter
BH configurations.

In the Kerr-de Sitter BH background, we
have observed that with the increase of the
cosmological constant, the superradiant
instability becomes harder to be triggered. It
needs to put the matter shell closer to the hole
and increase the coupling between the matter and
scalar field to spark off the instability. We
have got the physical understanding of the
phenomenon we observed by examining the potential
behavior.

Further, we have examined the influence of the
angular momentum per unit mass on the
superradiant instability.  We have disclosed that
slower rotation will make the superradiant
instability harder to happen. There is a minimum
value of the angular momentum per unit mass to
allow the appearance of the superradiant
instability. This minimum $a$ increases when the
cosmological constant becomes bigger. The
physical reasons behind this phenomenon have also
been explained.

On the superradiant instability study, the
difference between asymptotically flat Kerr black
hole and Kerr-de Sitter BH is very
little. The main reason is that the superradiant
instability is mainly influenced by the potential
barrier to reflect back the outgoing perturbation
and the potential well between the BH
horizon and the potential barrier to store the
reflected perturbation. It has little to do with
the boundary condition at the infinity for the
asymptotically flat spacetime or at the
cosmological horizon for the de Sitter spacetime.

We have also examined the spontaneous
scalarization in the de Sitter BH
background. Different from the superradiant
instability, now the boundary at the cosmological
horizon becomes important in the investigation.
Different from the asymptotically flat BH
case \cite{Cardoso}, the regularity condition at
the cosmological horizon enforces the scalarization to take a special style, that is the
scalar field takes a constant profile outside the shell as well as inside the shell in the spherical de Sitter BH background.  Since we only focused on the
spherical de Sitter configurations, one question
we may ask is that whether a more general scalarization can
happen once the de Sitter BH background
starts to rotate. Further investigations on this
topic are called for.

\section*{Acknowledgments}

We thank De-Chang Dai, Xiao-Mei Kuang, De-Cheng
Zou, Zhi-Ying Zhu and Yun-Qi Liu for their
helpful suggestions and discussions. This work
was supported in part by the National Natural Science
Foundation of China.


\begin{thebibliography}{99}

\bibitem{BD}C. H. Brans and R. H. Dicke, Phys. Rev. {\bf 124}, 925 (1961).

\bibitem{STBook}
Y.~Fujii and K.~Maeda,
  %``The scalar-tensor theory of gravitation,''
  Cambridge, USA: Univ. Pr. (2003) 240 p\\
  %17 citations counted in INSPIRE as of 14 May 2014
 V.~Faraoni, Springer (2004) 274 p
  %``Cosmology in scalar tensor gravity,''
  %ISBN-1-4020-1988-2.
  %%CITATION = ISBN-1-4020-1988-2;%%
  %5 citations counted in INSPIRE as of 14 May 2014

%\cite{Green:1987mn}
\bibitem{Green:1987mn}
  M.~B.~Green, J.~H.~Schwarz and E.~Witten,
  %``Superstring Theory. Vol. 2: Loop Amplitudes, Anomalies And Phenomenology,''
  Cambridge, Uk: Univ. Pr. (1987) 596 P. (Cambridge Monographs On Mathematical Physics)
  %40 citations counted in INSPIRE as of 14 May 2014

\bibitem{test}J.~Khoury and A.~Weltman,
  %``Chameleon fields: Awaiting surprises for tests of gravity in space,''
  Phys.\ Rev.\ Lett.\  {\bf 93}, 171104 (2004)
  [arXiv:astro-ph/0309300].\\
 K.~Hinterbichler and J.~Khoury,
  %``Symmetron Fields: Screening Long-Range Forces Through Local Symmetry Restoration,''
  Phys.\ Rev.\ Lett.\  {\bf 104}, 231301 (2010)
  [arXiv:1001.4525 [hep-th]].

\bibitem{hair}T. Damour and G. Esposito-Farese, Phys. Rev. Lett.
{\bf 70}, 2220 (1993).\\T.~Damour and G.~Esposito-Farese,
  %``Tensor - scalar gravity and binary pulsar experiments,''
  Phys.\ Rev.\ D {\bf 54}, 1474 (1996)
  [arXiv:gr-qc/9602056].\\ P.~Pani, V.~Cardoso, E.~Berti, J.~Read and M.~Salgado,
  %``The vacuum revealed: the final state of vacuum instabilities in compact stars,''
  Phys.\ Rev.\ D {\bf 83}, 081501 (2011)
  [arXiv:1012.1343 [gr-qc]].\\
   E.~Barausse, C.~Palenzuela, M.~Ponce and L.~Lehner,
  %``Neutron-star mergers in scalar-tensor theories of gravity,''
  Phys.\ Rev.\ D {\bf 87}, 081506 (2013)
  [arXiv:1212.5053 [gr-qc]].\\
   C.~Palenzuela, E.~Barausse, M.~Ponce and L.~Lehner,
  %``Dynamical scalarization of neutron stars in scalar-tensor gravity theories,''
  Phys.\ Rev.\ D {\bf 89}, 044024 (2014)
  [arXiv:1310.4481 [gr-qc]].



\bibitem{Uniq}S.W. Hawking, Comm. Math. Phys. {\bf 25}, 167 (1972).\\ T.~P.~Sotiriou and V.~Faraoni,
  %``Black holes in scalar-tensor gravity,''
  Phys.\ Rev.\ Lett.\  {\bf 108}, 081103 (2012)
  [arXiv:1109.6324 [gr-qc]].

\bibitem{different} V.~Cardoso, S.~Chakrabarti, P.~Pani, E.~Berti and L.~Gualtieri,
  %``Floating and sinking: The Imprint of massive scalars around rotating black holes,''
  Phys.\ Rev.\ Lett.\  {\bf 107}, 241101 (2011)
  [arXiv:1109.6021 [gr-qc]].\\ N.~Yunes, P.~Pani and V.~Cardoso,
  %``Gravitational Waves from Quasicircular Extreme Mass-Ratio Inspirals as Probes of Scalar-Tensor Theories,''
  Phys.\ Rev.\ D {\bf 85}, 102003 (2012)
  [arXiv:1112.3351 [gr-qc]].


\bibitem{Cardoso}V.~Cardoso, I.~P.~Carucci, P.~Pani and T.~P.~Sotiriou,
  %``Matter around Kerr black holes in scalar-tensor theories: scalarization and superradiant instability,''
  Phys.\ Rev.\ D {\bf 88}, 044056 (2013)
  [arXiv:1305.6936 [gr-qc]].\\ V.~Cardoso, I.~P.~Carucci, P.~Pani and T.~P.~Sotiriou,
  %``Black holes with surrounding matter in scalar-tensor theories,''
  Phys.\ Rev.\ Lett.\  {\bf 111}, 111101 (2013)
  [arXiv:1308.6587 [gr-qc]].

\bibitem{Weinberg:2008zzc}
  S.~Weinberg,
  %``Cosmology,''
  Oxford, UK: Oxford Univ. Pr. (2008) 593 p
  %37 citations counted in INSPIRE as of 14 May 2014

\bibitem{Hod2013}
S.~Hod,
  %``Stability of the extremal Reissner-Nordstroem black hole to charged scalar perturbations,''
  Phys.\ Lett.\ B {\bf 713}, 505 (2012)
  [arXiv:1304.6474 [gr-qc]].


\bibitem{Cardoso2013}
 V.~Cardoso,
  %``Black hole bombs and explosions: from astrophysics to particle physics,''
  Gen.\ Rel.\ Grav.\  {\bf 45}, 2079 (2013)
  [arXiv:1307.0038 [gr-qc]].

\bibitem{QNM of Kerr-dS}S. Yoshida, N. Uchikata, T. Futamase, Phys.
Rev. D {\bf 81}, 044005 (2010).

\bibitem{redefine-scalar} T.~PSotiriou, V.~Faraoni and S.~Liberati,
  %``Theory of gravitation theories: A No-progress report,''
  Int.\ J.\ Mod.\ Phys.\ D {\bf 17}, 399 (2008)
  [arXiv:0707.2748 [gr-qc]].

\bibitem{alpha3} T.~Damour and G.~Esposito-Farese,
  %``Gravitational wave versus binary - pulsar tests of strong field gravity,''
  Phys.\ Rev.\ D {\bf 58}, 042001 (1998)
  [arXiv:gr-qc/9803031].\\ P.~C.~C.~Freire, N.~Wex, G.~Esposito-Farese, J.~P.~W.~Verbiest, M.~Bailes, B.~A.~Jacoby, M.~Kramer and I.~H.~Stairs {\it et al.},
  %``The relativistic pulsar-white dwarf binary PSR J1738+0333 II. The most stringent test of scalar-tensor gravity,''
  Mon.\ Not.\ Roy.\ Astron.\ Soc.\  {\bf 423}, 3328 (2012)
  [arXiv:1205.1450 [astro-ph.GA]].

\bibitem{spheroidal-Teukolsky}S. A. Teukolsky, Astrophys. J. {\bf 185},
635 (1973).

\bibitem{spheroidal-Seidel}E. D. Fackerell, R. G. Crossman, J. Math.
Phys. {\bf 18}, 1849 (1977).\\
E. Seidel, Class. Quant. Grav. {\bf 6}, 1057 (1989).

\bibitem{Levear-spheroidal}E. W. Leaver, Proc. R. Soc. London {\bf A402}, 285 (1985). \\
E. W. Leaver, Phys.\ Rev.\ D {\bf 34}, 384 (1986); \\
E. W. Leaver, J. Math.\ Phys.\ {\bf 27}, 1238 (1986).

\bibitem{Separation constant 2}H. Suzuki, E. Takasugi, and H. Umetsu,
Prog. Theor. Phys. {\bf 102}, 253 (1999).

\bibitem{Kerr-sep-const} E.~Berti, V.~Cardoso and M.~Casals,
  %``Eigenvalues and eigenfunctions of spin-weighted spheroidal harmonics in four and higher dimensions,''
  Phys.\ Rev.\ D {\bf 73}, 024013 (2006)
  [Erratum-ibid.\ D {\bf 73}, 109902 (2006)]
  [arXiv:gr-qc/0511111].

\bibitem{Separation constant} H.~Suzuki, E.~Takasugi and H.~Umetsu,
  %``Perturbations of Kerr-de Sitter black hole and Heun's equations,''
  Prog.\ Theor.\ Phys.\  {\bf 100}, 491 (1998)
  [arXiv:gr-qc/9805064].\\ H.~T.~Cho, A.~S.~Cornell, J.~Doukas and W.~Naylor,
  %``Asymptotic iteration method for spheroidal harmonics of higher-dimensional Kerr-(A)dS black holes,''
  Phys.\ Rev.\ D {\bf 80}, 064022 (2009)
  [arXiv:0904.1867 [gr-qc]].

%\cite{Brady:1999wd}
\bibitem{Brady:1999wd}
  P.~R.~Brady, C.~M.~Chambers, W.~G.~Laarakkers and E.~Poisson,
  %``Radiative falloff in Schwarzschild-de Sitter space-time,''
  Phys.\ Rev.\ D {\bf 60}, 064003 (1999)
  [arXiv:gr-qc/9902010].
  %%CITATION = GR-QC/9902010;%%
  %71 citations counted in INSPIRE as of 14 May 2014

%\cite{Molina:2003dc}
\bibitem{Molina:2003dc}
  C.~Molina, D.~Giugno, E.~Abdalla and A.~Saa,
  %``Field propagation in de Sitter black holes,''
  Phys.\ Rev.\ D {\bf 69}, 104013 (2004)
  [arXiv:gr-qc/0309079].
  %%CITATION = GR-QC/0309079;%%
  %44 citations counted in INSPIRE as of 14 May 2014


\bibitem{QNM-methods} E.~Berti, V.~Cardoso and A.~O.~Starinets,
  %``Quasinormal modes of black holes and black branes,''
  Class.\ Quant.\ Grav.\  {\bf 26}, 163001 (2009)
  [arXiv:0905.2975 [gr-qc]].\\ R.~A.~Konoplya and A.~Zhidenko,
  %``Quasinormal modes of black holes: From astrophysics to string theory,''
  Rev.\ Mod.\ Phys.\  {\bf 83}, 793 (2011)
  [arXiv:1102.4014 [gr-qc]].



\bibitem{P-Teller} P\"{o}schl, G. and Teller, E., Z. Phys. {\bf 83}, 143 (1933).\\V.~Cardoso and J.~P.~S.~Lemos,
  %``Quasinormal modes of the near extremal Schwarzschild-de Sitter black hole,''
  Phys.\ Rev.\ D {\bf 67}, 084020 (2003)
  [arXiv:gr-qc/0301078].\\V.~Cardoso, M.~Lemos and M.~Marques,
  %``On the instability of Reissner-Nordstrom black holes in de Sitter backgrounds,''
  Phys.\ Rev.\ D {\bf 80}, 127502 (2009)
  [arXiv:1001.0019 [gr-qc]].

\bibitem{Finite-diff}C.~Gundlach, R.~H.~Price and J.~Pullin,
  %``Late time behavior of stellar collapse and explosions: 1. Linearized perturbations,''
  Phys.\ Rev.\ D {\bf 49}, 883 (1994)
  [arXiv:gr-qc/9307009].\\C.~Gundlach, R.~H.~Price and J.~Pullin,
  %``Late time behavior of stellar collapse and explosions: 2. Nonlinear evolution,''
  Phys.\ Rev.\ D {\bf 49}, 890 (1994)
  [arXiv:gr-qc/9307010].\\R.~A.~Konoplya and A.~Zhidenko,
  %``Instability of higher dimensional charged black holes in the de-Sitter world,''
  Phys.\ Rev.\ Lett.\  {\bf 103}, 161101 (2009)
  [arXiv:0809.2822 [hep-th]].

\bibitem{WKB}G. Wentzel, Z. Physik {\bf 38}, 518 (1926); K.A. Kramers Z.
Physik {\bf 39}, 828 (1926); L. Brillouin, Compt. Rend. {\bf 183}, 24 (1926).

\bibitem{Directly Integ}S. Chandrasekhar, S. Detweiler, Proc.\ R.\
Soc.\ Lond. {\bf A344}, 441-452 (1975).\\C.~Molina, P.~Pani, V.~Cardoso and L.~Gualtieri,
  %``Gravitational signature of Schwarzschild black holes in dynamical Chern-Simons gravity,''
  Phys.\ Rev.\ D {\bf 81}, 124021 (2010)
  [arXiv:1004.4007 [gr-qc]].

\bibitem{Khanal1985}
U. Khanal, Phys.\ Rev.\ D {\bf 32}, 879 (1985);\\
T. Tachizawa and K.-ichi Maeda, Phys. Lett.\ A {\bf 172}, 325 (1993).

\bibitem{freque-decrease}I.~G.~Moss and J.~P.~Norman,
  %``Gravitational quasinormal modes for anti-de Sitter black holes,''
  Class.\ Quant.\ Grav.\  {\bf 19}, 2323 (2002)
  [arXiv:gr-qc/0201016].\\ S.~Yoshida and T.~Futamase,
  %``Numerical analysis of quasinormal modes in nearly extremal Schwarzschild-de Sitter space-times,''
  Phys.\ Rev.\ D {\bf 69}, 064025 (2004)
  [arXiv:gr-qc/0308077].\\A.~Maassen van den Brink,
  %``Approach to the extremal limit of the Schwarzschild-de sitter black hole,''
  Phys.\ Rev.\ D {\bf 68}, 047501 (2003)
  [arXiv:gr-qc/0304092].

%\bibitem{bomb}W. H. Press and S. A. Teukolsky, Nature {\bf 238}, 211-212
%(1972).\\V.~Cardoso, O.~J.~C.~Dias, J.~P.~S.~Lemos and S.~Yoshida,
%  %``The Black hole bomb and superradiant instabilities,''
%  Phys.\ Rev.\ D {\bf 70}, 044039 (2004)
%  [Erratum-ibid.\ D {\bf 70}, 049903 (2004)]
%  [hep-th/0404096].\\ P.~Pani, V.~Cardoso, L.~Gualtieri, E.~Berti and A.~Ishibashi,
%  %``Black hole bombs and photon mass bounds,''
%  Phys.\ Rev.\ Lett.\  {\bf 109}, 131102 (2012)
%  [arXiv:1209.0465 [gr-qc]].

\bibitem{Sufficient}W. F. Buell and B. A. Shadwick, Am.\ J.\ Phys.
{\bf 63}, 256 (1995).


\bibitem{potential} V.~Cardoso and S.~Yoshida,
  %``Superradiant instabilities of rotating black branes and strings,''
  JHEP {\bf 0507}, 009 (2005)
  [arXiv:hep-th/0502206].

\bibitem{Israel:1966rt}
  W.~Israel,
  %``Singular hypersurfaces and thin shells in general relativity,''
  Nuovo Cim.\ B {\bf 44S10}, 1 (1966)
  [Erratum-ibid.\ B {\bf 48}, 463 (1967)]
  [Nuovo Cim.\ B {\bf 44}, 1 (1966)].
  %%CITATION = NUCIA,B44S10,1;%%
  %1248 citations counted in INSPIRE as of 01 Jul 2014

\end{thebibliography}
\end{document}